\documentclass[a4paper, 11pt]{article}
\usepackage{latexsym,amsmath,amsfonts,amssymb}
\usepackage{tikz}
\usetikzlibrary{decorations.pathmorphing,cd,decorations.markings,calc}
\usepackage{mathrsfs}
\usepackage[american]{babel}
\usepackage{graphicx}
\usepackage{bbm}
\usepackage{cite}
\usepackage{tcolorbox}
\usepackage{cancel}

\usepackage[colorlinks=true, citecolor=blue!90!black, linkcolor=blue!90!black, linktocpage=true, urlcolor=red!70!black]{hyperref}

\renewcommand{\baselinestretch}{1.2}
\numberwithin{equation}{section}

\newcommand{\be}{\begin{equation}} \newcommand{\ee}{\end{equation}}
\newcommand{\bea}{\begin{equation} \begin{aligned}} \newcommand{\eea}{\end{aligned} \end{equation}}

\newcommand{\Z}{\mathbb{Z}}
\newcommand{\R}{\mathbb{R}}

\newcommand{\cC}{\mathcal{C}}
\newcommand{\cD}{\mathcal{D}}

\newcommand{\bQ}{\mathbb{Q}}
\newcommand{\bR}{\mathbb{R}}

\newcommand{\bZ}{\mathbb{Z}}

\def\repa{\raise4pt\hbox{$\square$}\mkern-14mu\raise-4pt\hbox{$\square$}}
\def\repab{\overline{\raise4pt\hbox{$\square$}\mkern-14mu\raise-4pt\hbox{$\square$}\mkern-1mu}}

\begin{document}

\thispagestyle{empty}
\fontsize{12pt}{20pt}
\vspace{13mm}
\begin{center}
	{\huge 
    Lattice Realizations of Flat Gauging and \\ [8pt] T-duality Defects at Any Radius
	\\[8mm]}
    	{\large Riccardo Argurio, Giovanni Galati and Nathan Godechal

}
	
	\bigskip
	{\it
		Physique Th\'eorique et Math\'ematique and International Solvay Institutes\\
Universit\'e Libre de Bruxelles, C.P. 231, 1050 Brussels, Belgium 
	}
\end{center}

\bigskip

\begin{abstract}

\noindent
We analyze non-invertible topological interfaces and defects in the two-dimensional compact boson, focusing on the more exotic ones obtained by gauging continuous symmetries with flat connections on a half-space. These include interfaces between mutually irrational radii and T-duality symmetries at arbitrary boson radius. Using the modified Villain discretization on both a Euclidean two-dimensional square lattice and a quantum one-dimensional chain, we show that all these topological interfaces survive discretization and give rise to non-compact edge modes localized at the defect sites. Such non-compact edge modes imply a continuous defect spectrum and an infinite quantum dimension. In the special case of rational radii, we show how the defect action or Hamiltonian can be modified in order to compactify the edge modes and produce more standard defects with finite quantum dimension.

\end{abstract}

\newpage
\pagenumbering{arabic}
\setcounter{page}{1}
\setcounter{footnote}{0}
\renewcommand{\thefootnote}{\arabic{footnote}}

{\renewcommand{\baselinestretch}{.88} \parskip=0pt
\setcounter{tocdepth}{2}
\tableofcontents}


\section{Introduction}

Topological manipulations in quantum systems can be understood as maps between theories that share the same local physics but may differ in global properties. In particular, they do not alter the RG flow. Familiar examples are orbifolds in two-dimensional conformal field theories, as well as discrete gaugings of higher-form symmetries \cite{Gaiotto:2014kfa}. In fact, one can more generally associate a topological interface with such a manipulation, viewed as an object separating the two corresponding theories \cite{Diatlyk:2023fwf}. When the topological manipulation maps a given theory to itself, it then implements a symmetry of the theory, and the interface becomes a symmetry defect (generically of the non-invertible kind). See \cite{McGreevy:2022oyu,Shao:2023gho,Schafer-Nameki:2023jdn} for recent reviews on generalized symmetries.

This perspective has recently been extended to include gaugings of continuous symmetries in which the gauge field is integrated only over flat connections, a manipulation dubbed \emph{flat-gauging} \cite{Brennan:2024fgj,Antinucci:2024zjp,Bonetti:2024cjk, Arbalestrier:2024oqg, Argurio:2024ewp,Paznokas:2025auw,Arbalestrier:2025poq,Oguz:2025ftx,Apruzzi:2025hvs,Arbalestrier:2025jsg}.\footnote{Actually, in the 2d CFT context, these manipulations were introduced some time ago and dubbed continuous orbifolds \cite{Gaberdiel:2011aa}, see also \cite{Fredenhagen:2012bw,Gaberdiel:2014vca}.} Such flat-gaugings lead to new consequences in some QFTs. In the case of the two-dimensional free compact boson, they imply the existence of a non-invertible T-duality defect at any irrational radius $R \in \bR$ and, more generally, allow one to relate theories at radii $R$ and $R'$ with arbitrary real ratio $R/R' \in \mathbb{R}$ \cite{Argurio:2024ewp}. While the corresponding topological defects/interfaces can be constructed, they display non-trivial features, in particular the emergence of infinite quantum dimensions, i.e.~infinite vacuum expectation values when considered on homologically trivial loops.
This is to be contrasted with the more standard situations, for instance the compact boson at the special values $R^2 \in \bQ$ \cite{Thorngren:2021yso}, where the non-invertible T-duality symmetry can be constructed using only gaugings of finite subgroups of the global symmetry, and the corresponding defect has finite quantum dimension.\footnote{See \cite{Chang:2018iay,Nagoya:2023zky,Damia:2024xju,Choi:2023vgk,Bharadwaj:2024gpj,Furuta:2024vtc,Arias-Tamargo:2025xdd} for further studies on duality symmetries in compact bosons theories.}

The aim of this paper is to analyze these interfaces in the two-dimensional compact boson from different angles.\footnote{The theory of a single compact boson provides the simplest set-up where these interfaces arise. The techniques developed in this work can be naturally extended to more general systems where similar physics arises, such as two-dimensional non-linear sigma models. This set-up has been recently studied in the continuum \cite{Arias-Tamargo:2025xdd,Arias-Tamargo:2025fhv} and on the lattice \cite{Bacq:2026ega}.} We first review in Section \ref{sec:continuum} the framework of \cite{Argurio:2024ewp} for the 2d CFT of a compact boson in a continuous spacetime, and we derive explicitly the one-dimensional theory supported on the non-invertible T-duality symmetry defect, including its non-trivial coupling to (both sides of) the bulk. This defect theory resolves the otherwise problematic quantization of the naive topological BF-like couplings at irrational radius by exhibiting a non-compact defect mode (i.e.~a non compact scalar living on the worldline of the defect), which is naturally tied to the flat gauging of the $U(1)$ symmetries of the compact boson. Its zero mode is also the source of the infinite quantum dimension of such defect.

We next analyze lattice regularizations of this theory. The main goal is to show that the existence of these topological interfaces is universally dictated by the symmetry content of the model, and that the appearance of non-compact edge modes is in fact generic.\footnote{Moreover, the study of generalized symmetries and their lattice realizations has revealed very interesting features, see e.g.~\cite{Cheng:2022sgb,Seifnashri:2023dpa,Seiberg:2023cdc,Pace:2024oys,Seiberg:2024gek}.} To this aim, in Section \ref{sec:euclideanlattice} we first consider the compact scalar model formulated on a two-dimensional Euclidean lattice. It is by now well-understood that in order to obtain a theory that has both shift and winding symmetries, as the continuum theory, one has to adopt the modified Villain prescription to avoid the presence of dynamical vortices \cite{Sulejmanpasic:2019ytl,Gorantla:2021svj}. We argue that such a prescription exactly mirrors the topological manipulation that one would perform in the continuum to go from a non-compact to a compact scalar, namely the (flat) gauging of a $\mathbb{Z}$ subgroup of the $\mathbb{R}$ shift symmetry of the non-compact scalar. Armed with this understanding, we show how to express quite directly a generic topological interface between two lattice theories at radii $R$ and $R'$, and we analyze in detail the non-invertible T-duality defect present in these theories. It can be seen as the (one-dimensional) lattice version of the continuum defect, including the non-compact mode. We then show how, in the fine-tuned case $R^2 \in \bQ$, one can obtain from this construction the more familiar T-duality defect, originally presented in \cite{Choi:2021kmx}, with only compact edge-modes.

Finally, in Section \ref{sec:hamiltonianlattice} we consider again the same theory, but on a spatial lattice, namely a chain \cite{Fazza:2022fss,Cheng:2022sgb}. In this context the implementation of the modified Villain prescription is quite natural. Most of the physics of the model is encoded in the gauge constraints (also called Gauss laws) at every site and link of the chain. Accordingly, we show how to locally modify the Hamiltonian and the Gauss laws to construct topological interfaces that connect different compact boson theories, as well as those that implement the non-invertible T-duality at any radius. Once again we find non-compact edge modes, which become compact in the particular case of $R^2 \in \bQ$. The infinite quantum dimension of the defect with non-compact edge modes translates in this set-up into a (possibly more mundane) continuous spectrum for the defect Hamiltonian.

\section{The Compact Boson and its Topological Interfaces}\label{sec:continuum}
In this section we briefly review the theory of a compact boson in the continuum, its shift and winding symmetries, and how one can implement a discrete symmetry by a chain of T-duality and a rescaling of the radius. In particular, at any given radius $R$, including irrational values of $R^2$, the latter operation can always be performed by a decompactification followed by a compactification at the desired new radius. Decompactification is achieved by a flat gauging of the $U(1)$ winding symmetry, while compactification is achieved by gauging a $\bZ$ subgroup of the $\bR$ shift symmetry. As we are going to show, the very same steps allow us to construct generic topological interfaces separating any two compact boson theories with generic radii $R,R'\in \bR$.\footnote{Moreover, upon fusing these interfaces with the known topological interfaces relating the circle and orbifold branches of the $c=1$ conformal manifold \cite{Fuchs:2007tx,Bachas:2007td}, one obtains a topological interface between any two $c=1$ CFTs. This construction also admits a generalization to Abelian double-current deformations interpolating between two 2d CFTs \cite{Dubovsky:2023lza}. It would be interesting to determine whether any two 2d CFTs connected by an exactly marginal deformation can, in turn, be related through a (flat) orbifold.}

Consider the compact boson action
\be
    S = \frac{R^2}{4 \pi} \int d\phi \wedge * d\phi\ ,
\ee
with $\phi\sim\phi+2\pi$. More precisely, we have that
$\phi \in C^{\infty}(M,S^1)$ such that $d\phi \in Z^1(M,2\pi\Z)$, a closed 1-form with integral periods. We can further use a Hodge decomposition for $d\phi$
\be
    d\phi = d\phi^\R +\sum_{i=1}^{b_1} 2\pi n_i\hat{\gamma_i}\ ,
\ee
with  $\phi^\R \in C^\infty(M,\R)$, $\hat\gamma_i$ a basis of $H^1(M,\bZ)$, $n_i \in \Z $ and $b_1 = \dim(H^1(M,\bZ))$. The different values for the $n_i$'s divide the configuration space into distinct topological sectors.

We recall now the symmetries of the model. The conserved current for the shift, or momentum, symmetry is obtained from the equations of motion
\be
    j^m = -i\frac{R^2}{2\pi}d\phi\ .
\ee
Since the current is obtained from the equations of motion, the associated symmetry acts explicitly on the fields of the Lagrangian according to
\be \label{SYM:MOM}
    \phi \rightarrow \phi + \lambda\ ,
\ee
with $d\lambda = 0$, that is $\lambda$ is a constant, assuming $M$ is connected. This is the action of the global momentum/shift  symmetry on the field. The topological line operator implementing the symmetry is 
\be 
    U^m_\lambda (\Sigma_1) = \exp \left( i\lambda \int_{\Sigma_1} *j^m \right) \ ,
\ee
with $\lambda \in U(1)$ and $\Sigma_1$ closed, while the local
operators charged under it are the vertex operators $e^{in\phi}$ with $n\in \bZ$.

The conserved current for the winding symmetry is obtained from the flatness of $d\phi$
\be
j^w = \frac{1}{2\pi}*d\phi\ .
\ee
The winding symmetry probes the topological sector of the $S^1$ valued field $\phi$. Indeed, the topological line operator for the winding symmetry is
\be 
     U^w_{\lambda'}(\Sigma_1)  = \exp \left(\frac{i \lambda'}{2\pi} \int_{\Sigma_1}d\phi \right) \ ,
\ee
again with $\lambda' \in U(1)$ and $\Sigma_1$ closed.
The local operators charged under it are built as follows. Excise one point $x$ that is encircled by $\Sigma_1$, so that $M$ subtracted by this point has a non-trivial cycle, which coincides with $\Sigma_1$. 
We can then impose the path integral to be over configurations such that 
\be 
\int_{\Sigma_1}d\phi = 2\pi n\ .
\ee
Over such configurations, we have
\be
    \langle U^w_{\lambda'}(\Sigma_1) \rangle =  e^{i\lambda' n}\ .
\ee
Such a local operator, charged under the winding symmetry, can be thought as a vortex, or equivalently as a disorder operator for winding. Let us finally mention that the momentum and winding symmetries have a mixed anomaly, essentially due to the fact that both currents are expressed in terms of $d\phi$.

Anticipating what we are going to do on the lattice, we can have a first intuitive look at the relation between gauging the winding symmetry and decompatification.
Gauging entirely the winding symmetry amounts to imposing
\be
   \langle U^w_{\lambda'}(\Sigma_1) \rangle = 1 \ ,
\ee
for any $\Sigma_1\subset M$. 
As a result, we loose our ability to detect the winding of the boson around a vortex, so that the latter are no longer in the spectrum of genuine local operators. On the other hand, local operators non-integrally charged under the momentum symmetry such as $e^{ip\phi}$, with $p\in \bR$, are now genuine because the lines to which they were attached have become transparent.
This is the spectrum of a non-compact scalar. Note that the gauging procedure we have used mimicks the one for discrete symmetries, in particular we are not introducing degrees of freedom from the gauging. This translates to the fact that we have effectively gauged the winding $U(1)$ symmetry with flat connections.

More generally, what we mean by gauging a symmetry $G$ of a theory $S[\phi]$ with flat connections, is defining a new theory in which the partition function includes an additional sum over all the symmetry operator insertions. Let us consider a $p$-form symmetry with symmetry group $G$, the gauging of this symmetry amounts to considering the partition function
\be
    Z' =  \frac{1}{|G| \dim H_{d-p-1}} \sum_{[\Sigma_{d-p-1}] \in H_{d-p-1}(M)}\int_G d\alpha\int \cD \phi  \exp \left( -S[\phi]\right) U_\alpha (\Sigma_{d-p-1})\ .
\ee
Since $U_\alpha (\Sigma_{d-p-1})$ can be written as 
\be
U_\alpha (\Sigma_{d-p-1})=\exp\left(i\alpha\int_M *j\wedge \hat{\Sigma}_{d-p-1}\right)\ , 
\ee
with $\partial \Sigma_{d-p-1}=0$ implying $d\hat{\Sigma}_{d-p-1}=0$, 
the above procedure is 
equivalent to gauging the symmetry by choosing a connection defined on a principal $G$-bundle over $M$, but restricting the possible connections to flat ones.

Therefore, at the level of the action, in order to gauge the winding symmetry we couple the winding current to a $U(1)$ gauge field $c$. Then we add a third term enforcing the flatness of the gauge field $c$. We end up with the action 
\begin{equation} \label{XY:GAUGE_WIND}
S=\frac{R^2}{4\pi} \int d\phi \wedge *d \phi + \frac{i}{2\pi}  \int c\wedge d\phi  - \frac{i}{2\pi R} \int c\wedge d\psi^\R \ ,
\end{equation}
with $\psi^\R$ valued in $\R$. The normalization of the third term is chosen for convenience since $\psi^\R$ is a $\R$ gauge field and can thus be freely normalized. We then integrate out $c$ which enforces through a delta function the relation:
\begin{equation}
    d\phi = \frac{1}{R}d\psi^\R\ .
\end{equation}
We end up with the action:
\begin{equation}
    S = \frac{1}{4\pi} \int d\psi^\R \wedge *d\psi^\R\ ,
\end{equation}
which is the action for a non-compact boson.

Let us now conversely gauge a $\bZ$ subgroup of the $\bR$ shift/momentum symmetry of the non-compact boson above. We have to gauge the shift symmetry by an $\R$ connection $b$, and then impose that the latter has holonomies in $\Z$ by adding a coupling to a compact scalar $\phi'$
\be
S = \frac{1}{4\pi} \int (d\psi^\R - b)\wedge *(d\psi^\R - b) + \frac{iR'}{2\pi} \int  b \wedge d\phi'\ .
\ee
The constant $R' \in \bR$ parametrizes all possible embeddings of $\bZ \subset \bR$.

The equations of motion for $b$ lead to 
\be
*(d\psi^\R-b)=iR'd\phi'\ ,
\ee
so that substituting back into the action yields
\be
S=\frac{R'^2}{4\pi} \int d\phi' \wedge *d \phi' +\frac{iR'}{2\pi}\int d\psi^\R \wedge d \phi'\ ,
\ee
describing a compact boson at radius $R'$.
The second term in the action above can be safely integrated to zero on a compact manifold. We however retain it here since it plays an important role precisely when the gauging is performed on a manifold with a boundary (namely, only on part of spacetime). Another important remark to be done is that in this way of gauging the momentum symmetry, we are in fact also performing a T-duality. More specifically, since the relation between the initial and the final variables, $\psi^\R$ and $\phi'$ respectively, involves a $*$, it means that momentum and winding symmetries are actually exchanged in the process.\footnote{As noted e.g.~in \cite{Argurio:2024ewp}, T-duality for a compact scalar is implemented by gauging a (trivial) $\Z_1$ subgroup of its $U(1)$ momentum symmetry.}

\subsection{Topological Interfaces and the T-duality Defect}
We have now all the elements to build a topological interface connecting a compact boson at radius $R$ and the one at radius $R'$. We are basically going to do the two operations described above but only on half of spacetime. Among all these interfaces, we will then focus on the special one which connects the radius $R$ to itself in a non trivial way. This is the non-invertible T-duality defect.

In order to do that, we first insert a trivial interface between the left and right parts of the two dimensional spacetime. Then we perform the flat gauging of the whole winding symmetry on the right, to get a non-compact scalar there. Finally, we compactify again to radius $R'$ by gauging the appropriate $\bZ$ subgroup of the momentum $\bR$ symmetry, still on the right. As commented above, the latter operation also implements T-duality. These are all topological operations, hence the interface is itself topological. If $R' = R$, the final theory on the right being exactly the same as the one on the left, namely the one we started with, we conclude that such an interface is actually a symmetry defect. 

Let us then start from
\be
   S= S_L + S_R = \sum_i \frac{R^2}{4\pi} \int_{\Gamma_i} d\phi_i\wedge *d\phi_i \qquad i = L,R\qquad \partial \Gamma_L = -\partial \Gamma_R = I\ .
\ee
We can actually add the following action on the interface to enforce $\phi_L(x)|_{x\in I}=\phi_R(x)|_{x\in I}$ up to periodicity:
\be\label{trivialinterface}
S_I=\frac{i}{2\pi}\int_I (\phi_L-\phi_R) d\phi\ ,
\ee
with $\phi$ a $U(1)$-valued edge mode.\footnote{The path integral over the edge mode $\phi$ sets $\phi_L|_I=\phi_R|_L+2\pi\bZ$, while the boundary piece of the equations of motion for $\phi_L$ and $\phi_R$ together imply continuity of the derivative orthogonal to the interface. }
Then we flatly gauge the $U(1)$ winding symmetry on the right 
\be
    S_R \rightarrow \frac{R^2}{4\pi} \int_{\Gamma_R} d\phi_R \wedge *d\phi_R + 
\frac{i}{2\pi} \int_{\Gamma_R} c \wedge d\phi_R -\frac{i}{2\pi R} \int_{\Gamma_R} c \wedge d\psi^{\bR}\ ,
\ee
with $c$ a $U(1)$ connection and $\psi^\bR$ a real field. Integrating out $c$ gives $\phi_R=\frac{1}{R}\psi^\bR+2\pi \bZ$ and the action becomes
\be
S=\frac{R^2}{4\pi}\int_{\Gamma_L}d\phi_L\wedge *d\phi_L + \frac{1}{4\pi}\int_{\Gamma_R}d\psi^\bR\wedge *d\psi^\R +\frac{i}{2\pi}\int_I(\phi_L-\frac{1}{R}\psi^\bR) d\phi\ .
\ee
Eventually, we gauge a $\Z$ subgroup of the momentum symmetry of $\psi^\bR$:
\be
S_R \to \frac{1}{4\pi}\int_{\Gamma_R}(d\psi^\bR-b)\wedge *(d\psi^\R-b) +\frac{iR'}{2\pi}\int_{\Gamma_R}b\wedge d\phi'_R\ ,
\ee
with $b$ a real connection and $\phi'_R$ a $U(1)$ field (that makes $b$ a $\bZ$ gauge field). Integrating out $b$ we get
\be
S=\frac{R^2}{4\pi}\int_{\Gamma_L}d\phi_L\wedge *d\phi_L +S_{R| R'}+ \frac{R'^2}{4\pi}\int_{\Gamma_R}d\phi_R'\wedge *d\phi_R' \ ,
\ee
with
\be\label{tdualitydefectlong}
S_{R|R'} = \frac{i}{2\pi}\int_I(\phi_L-\frac{1}{R}\psi^\bR) d\phi+R'\psi^\bR  d\phi'_R\ .
\ee
The interface action as above is perfectly gauge invariant and manifestly topological. We observe that this interface is described by coupling two edge modes, $\phi$ and $\psi^{\mathbb{R}}$, to the bulk. Crucially, one of these modes is non-compact, i.e.\ $\mathbb{R}$-valued. This signals that the interface has infinite quantum dimension, as reflected in the divergence of its expectation value when evaluated on a circular loop.\footnote{Indeed, this expectation value reduces to computing the partition function of the one-dimensional TQFT on the defect, which in turn equates to the dimension of its Hilbert space. The non-compact defect TQFT is equivalent to a quantum particle on a ring ($\phi$ is the particle's position, and $\psi^\R$ its momentum), whose Hilbert space is infinite dimensional.} 

As mentioned before, in the special case $R=R'$ we actually get the non-invertible T-duality defect action $S_{R| R}\equiv S_T$. Let us then analyze some of the properties of this defect. If we were to integrate out carelessly $\phi$ or $\psi^\bR$, then we would get the badly quantized interface action $\frac{iR^2}{2\pi}\int_I \phi_L d\phi'_R$. Note that if we integrate out $\phi$, we could also choose to remain with the gauge invariant interface action 
\be\label{tdualitydefectshort}
S_T=\frac{iR}{2\pi}\int_I\psi^\bR  d\phi'_R \ ,
\ee
where one has to further set $\phi_L|_I=\frac{1}{R}\psi^\bR$ as an extra constraint in the path integral. This implies that the left compact boson $\phi_L$ is decompactified on the interface, and then it is coupled to the right degrees of freedom. This presentation of the defect will be useful in the lattice regularization of the theory.

Since the full bulk--defect action is still quadratic, we can easily derive the consequences of the T-duality defect by looking at the various equations of motion.
By examining the ones of the well defined action \eqref{tdualitydefectlong} (with $R'=R$)\footnote{Note that this is the most convenient presentation of the defect action for deriving the equations of motion, since all edge modes appear explicitly in the Lagrangian. In more implicit presentations, such as \eqref{tdualitydefectshort}, some equations of motion are not manifest, precisely because the edge mode $\phi$ is integrated out.} one readily finds that on the interface $I$ the following gluing conditions are valid
\begin{align}
d\phi'_R = \frac{1}{R^2}\, d\phi  \quad,\quad  d\phi_L = \frac{1}{R}\, d\psi^{\mathbb{R}} \quad,\quad
i* d\phi_L &=\,d\phi'_R  \, .
\end{align}
Therefore, for irrational values of $R^2$, the full set of $U(1)_m \times U(1)_w$ topological operators is trivialized by the defect, thus showing its non-invertibility. On the other hand, the defect acts invertibly on the two local operators $(d\phi_L,i*d\phi_L)$, mapping them to $(i*d\phi'_R,d\phi'_R)$. 

In the special case $R^2=\frac{p}{q}$, we have
\begin{equation}
\int_I d\phi'_R \;=\; \int_I i* d\phi_L \;\in\; \frac{2\pi q}{p}\,\mathbb{Z}\,.
\end{equation}
It follows that a non-trivial subset of the bulk topological operators survives. For these special values of $R$, one can in fact construct a minimal T-duality defect by gauging a $\mathbb{Z}_p \times \mathbb{Z}_q \subset U(1)_m \times U(1)_w$ symmetry in combination with T-duality.

In our framework, this defect emerges from the observation that, at these radii, the defect action can be rewritten as
\begin{equation}\label{eq:pq_Tdefect}
S_T\Big|_{R^2=\frac{p}{q}}
=
\frac{i}{2\pi}\int_I
(\phi_L-q\psi'^{\mathbb{R}})\,d\phi
+
p\psi'^{\mathbb{R}}\,d\phi'_R \, ,
\end{equation}
where
\begin{equation}
\psi'^{\mathbb{R}} := \frac{1}{\sqrt{pq}}\,\psi^{\mathbb{R}} \, .
\end{equation}
In this form, it is natural to promote the edge mode $\psi'^{\mathbb{R}}$ to a compact compact one, since the action remains gauge invariant.\footnote{Notice that the same arguments apply for the topological interfaces \eqref{tdualitydefectlong} in the special cases $R'R = \frac{p}{q}$.} One can then verify that this reproduces the standard T-duality defect, as written in e.g. \cite{Niro:2022ctq}.

This also shows that the defect described by the action \eqref{eq:pq_Tdefect} can be obtained from the minimal defect with compact $\psi$ by stacking it with a condensation defect for the winding symmetry, which effectively decompactifies the edge mode.

\section{Euclidean Lattice Regularization}
\label{sec:euclideanlattice}

We start this section by reviewing the problem of discretizing U(1) degrees of freedom and the solution which goes under the name of Modified Villain prescription.

Let us consider a QFT defined in the continuum, with a U(1) degree of freedom, meaning that some fields take values on a circle. Examples include a compact boson or a U(1) gauge field. In the following, we focus on a two-dimensional compact boson, as it already exhibits all the subtle features we wish to highlight, and it is the model we already discussed in the previous section. As we have already seen there, a key characteristic of this theory is the presence of topological sectors, namely field configurations that allow for distinct winding classes of the scalar field for which $\int d\phi \in 2\pi \mathbb{Z}$.
Since this integral yields discrete values, continuous deformations of the field configuration cannot alter it. 

When discretizing the space-time, we get a 2d lattice model,\footnote{Here we will take a square lattice $\Lambda$ consisting of vertices $v$, links $l$ between them, and (square) plaquettes $p$ enclosed by four links. The lattice can be taken to be toroidal for definiteness, though we will focus only on its local features.} with scalar fields valued on each site of the lattice. In this case the configuration space is
\be
\cC= \prod\limits_{v \in \Lambda} U(1)_v\,.
\ee
We see that, by discretizing the space-time, we get a simply connected configuration space, thus forbidding the splitting into superselection sectors. Indeed now, by trying to engineer a configuration such that $\sum_{v\in \gamma}d \phi\simeq\int_\gamma d\phi = 2\pi w \not=0$, we see that since $\phi_v\in\ ]-\pi,\pi]$, we eventually always undo the winding between the next-to-last and last lattice sites of the loop.

Therefore, in this naive discretization of the theory, topological sectors appear merely as emergent phenomena in the continuum limit. However, these features often play a crucial role in deriving fundamental constraints on the theory: they give rise to additional global symmetries and anomalies. This suggests that a more refined approach is necessary—one that goes beyond the naive discretization—to determine whether it is possible to construct a lattice model in which such subtleties manifest explicitly at the discrete level.

\subsection{Villain and Modified Villain Models}
One way to explicitly construct a lattice model exhibiting topological sectors is through the so-called Villain model \cite{Villain:1974ir,Berezinskii:1970pzv}. The basic idea of the Villain model is to see $U(1)$ as the quotient $\R/\Z$. Accordingly, one introduces a pair of degrees of freedom denoted $\phi_v$, valued in $\R$, and $n_l$, valued in $\Z$, respectively defined on vertices and links. Crucially, they have the following gauge transformation 
\be\label{eq: Z gauge transf}
\phi \rightarrow \phi + 2\pi k_\phi\quad,\quad n \rightarrow n + dk_\phi\,.
\ee
In this context, the winding numbers arise from the $\bZ$ variables $n$. 

However, the Villain model allows for dynamical vortices generated by $dn\not= 0$. If we think of the lattice as embedded in the continuum, the vortices come from the fact that each lattice plaquette can be considered as a hole in the continuum. However, such configuration are not present in the continuum theory we are aiming to describe, since the model does not contain dynamical vortices (and hence has winding symmetry). To achieve this property, we can explicitly suppress vortices by introducing a delta function $\delta(dn)$ or equivalently a Lagrange multiplier $\phi'$ which sets $dn=0$ in the path integral. Such modification leads to the so-called \emph{modified Villain model} \cite{Sulejmanpasic:2019ytl,Gorantla:2021svj} (see also \cite{Chen:2024ddr} for a nice review).

We can finally write the most accurate lattice theory describing a 2d free compact boson:
\be\label{eq: mvmodel}
Z = \left(\prod\limits_{v}\int_{-\pi}^{\pi} \frac{d\phi_v}{2\pi}\right)\left( \prod\limits_{l}\sum_{n_l\in\bZ} \right)\left(\prod_p \delta(dn_p)\right)\exp{\left(-\frac{\beta}{2}\sum_l(d\phi-2\pi n)_l^2\right)}\,.
\ee
Notice that in defining \eqref{eq: mvmodel}, we have implicitly chosen a gauge fixing which restricts the values of the non-compact boson $\phi$ to a set of representatives. More specifically, we have written $\phi\in\bR$ as $\phi + 2\pi k_\phi$, with $\phi\in\ ]-\pi,\pi]$ and $k_\phi\in\bZ$. The parameter $\beta$ appearing in the action can be related to the radius $R$ through the identification $\beta = \frac{R^2}{2\pi}$.

Intuitively, the fact that a U(1) scalar variable can be represented as a pair of vertex and link variables with a flatness condition can be understood as follows. To define a compact scalar in a 2d continuous space-time, we need to divide our manifold in patches $U_i$. In each patch the scalar is a well-defined 0-form---i.e.~it is a non compact scalar. In the double intersections we need to define transition functions $n_{ij}$ gluing together different representatives of the compact scalar. The Villain model is obtained by shrinking to zero the size of each patch: all the patches then become the sites of the lattice while the double intersections represents the links between sites. Therefore, the link variables are the analogue of the transitions functions. In this sense, the flatness condition $dn=0$ corresponds to the cocycle condition for the transition functions.

Finally, after introducing the gauge transformations \eqref{eq: Z gauge transf} in to partition function to leave $\phi$ unfixed, \eqref{eq: mvmodel} can be written as
\be\label{eq: mvmodel_}
Z = \frac{\left(\prod\limits_{v}\int_{-\infty}^{\infty} \frac{d\phi_v}{2\pi}\right)\left( \prod\limits_{l}\sum_{n_l\in\bZ} \right)\left(\prod_p \delta(dn_p)\right)\exp{\left(-\frac{\beta}{2}\sum_l(d\phi-2\pi n)_l^2\right)}}{\prod_v \text{Vol}(\bZ)_{(k_{\phi})_v}}\,,
\ee
where $\text{Vol}(\bZ)_{(k_{\phi})_v}$ denotes a choice of gauge fixing at vertex $v$. Instead of fixing $\phi$ to a $2\pi$ interval as above, we can also almost gauge fix $n$ to $0$ everywhere using the modified Villain constraint of closeness $dn=0$. But because the gauge parameter $k_\phi$ appears in an exact form in \eqref{eq: Z gauge transf}, we cannot gauge fix $n$ to $0$ along non-trivial $1$-cycles. Indeed, after a Hodge decomposition of the closed form $n$, and an absorption of the exact part into the gauge transformation \eqref{eq: Z gauge transf}, the partition function can be written as
\be\label{eq: mvmodel_2}
Z = \left(\prod\limits_{v}\int_{-\infty}^{\infty} \frac{d\phi_v}{2\pi}\right)\left( \prod_{i=1}^{b_1}\sum_{n_i\in\bZ} \right)\exp{\left(-\frac{\beta}{2}\sum_l(d\phi-2\pi n_i \hat{\gamma}_i)_l^2\right)}\,.
\ee
This presentation of the modified Villain model corresponds to a $\R$ boson with a sum of insertions of momentum defects along the non-trivial $1$-cycles $\gamma_i \in H^1(\Lambda)$ and makes more apparent the connection between the modified Villain lattice model and the continuum compact boson.

\subsection{Symmetries and Anomalies}
As nicely explained in \cite{Gorantla:2021svj}, the modified Villain model has the advantage to possess the same symmetries (and anomalies) of its continuum limit. The shift symmetry is manifest: the action is invariant under 
\be
\phi \rightarrow \phi + \lambda
\ee
with $\lambda \in Z^0(\Lambda,U(1))$, i.e.~a constant.  The fact that $\lambda$ is valued in U(1) follows from the gauge transformations \eqref{eq: Z gauge transf}. To see the winding symmetry more explicitly, we can rewrite 
\be
\prod_p \delta(dn_p)=\prod_p  \int_{-\pi}^{\pi} \frac{d\phi'_p}{2\pi}\exp{\left(i\sum_p \phi'_pdn_p\right)}\,.
\ee
Thus the winding symmetry acts as
\be
\phi' \rightarrow \phi' + \lambda'\ ,
\ee
again for a constant $\lambda'$. It is a symmetry because $\sum_p dn_p=0$ by the lattice version of the Stokes theorem. 
$ \phi'_p$ can be interpreted as the dual scalar. More precisely, the dual scalar is $\widetilde{\phi}'= * \phi'$ which is a 0-cochain living on the dual lattice\footnote{In the present two-dimensional context, the dual lattice is readily obtained by associating to every plaquette $p$ a dual vertex $\tilde v$, to every link $l$ a dual link $\tilde l$, and to every vertex $v$ a dual plaquette $\tilde p$. The operation $*$ maps fields defined on any original lattice element to dual fields defined on the corresponding element of the dual lattice.} and the shift symmetry acts on it as
\be
\widetilde{\phi}'\rightarrow \widetilde{\phi}' + \widetilde{\lambda}'\,,
\ee
where $\lambda'= *  \widetilde{\lambda}'$.

From the above transformation rules one can extract the form of the currents. In particular, the current for the momentum symmetry is obtained by varying the action with the momentum shift $\phi \rightarrow \phi + \lambda$ with $\lambda$ not closed\footnote{The conserved current on the lattice is defined analogously to the conserved current in the continuum, namely by $d*j=0$, or equivalenty $\delta j = 0$ (where $\delta=*d*$, see \cite{Sulejmanpasic:2019ytl}). Indeed, if the action is invariant under closed shifts $\phi \rightarrow \phi + \lambda$, the variation of the action under non-closed shifts is
\begin{equation}
    \delta S = i\sum_{c^{n+1}} (d\lambda)_{c^{n+1}} (j)_{c^{n+1}}= \pm i\sum_{c^n}\lambda_{c^n}(\delta j)_{c^n}=\pm i  \sum_{\tilde c^{d-n}}(* \lambda)_{\tilde c^{d-n}}(d* j)_{\tilde c^{d-n}}\ .
\end{equation}
Since this variation vanishes on the equation of motion, we indeed get $\delta j = 0$ after having integrated by parts.}
\begin{equation}
    \delta S_{\lambda} = \frac{\beta}{2}\sum_l \big((d\phi + d\lambda - 2\pi n)_l^2 - (d\phi - 2\pi n)_l^2\big) \approx \beta\sum_l  (d\lambda)_l (d\phi - 2\pi n)_l\,.
\end{equation}
The momentum current is thus $j^m = -i\beta (d\phi - 2\pi n)$. For the winding symmetry, we have:
\begin{equation}
    \delta_{\tilde{\lambda}'} S \approx i\sum_{p} (*\tilde{\lambda}')_p (dn)_p = i\sum_{\tilde{v}} (\tilde{\lambda}')_{\tilde{v}} (*dn)_{\tilde{v}} =  -i\sum_{\tilde{l}}(d\tilde{\lambda}')_{\tilde{l}} (*n)_{\tilde{l}}\ .
\end{equation}
The winding current is thus: $\tilde{j}^w = -*n$ and lives on the dual lattice.

Written like this, the winding current is not gauge invariant. However we can easily redefine it  as 
\be
\tilde{j}^w = \frac{1}{2\pi}*(d\phi - 2\pi n)\ ,
\ee
which is indeed conserved and gauge invariant. 

The conserved charges built out of these two currents are:
\begin{equation}
    \begin{aligned} \label{XY:SYM:Q}
       & Q^m(\tilde{\Sigma}_1) = \sum_{\tilde{l} \in \tilde{\Sigma}_1} (*j^m)_{\tilde{l}} = -i\beta \sum_{\tilde{l} \in \tilde{\Sigma}_1} *(d\phi - 2\pi n)_{\tilde{l}} \ ,\\
       & Q^w(\Sigma_1) = \sum_{l \in \Sigma_1} (*\tilde{j}^w)_l = \frac{1}{2\pi}\sum_{l \in \Sigma_1} (d\phi-2\pi n)_l\ .
    \end{aligned}
\end{equation}

These two charges are both topological thanks to the conservation of the local currents. We stress that $Q^m(\tilde{\Sigma}_1)$ is conserved on the equations of motions, i.e. for the field configurations such that $\delta j^m = 0$, while $Q^w(\Sigma)$ is conserved off-shell since $\delta \tilde{j}^w = 0$ holds for all field configurations in the modified Villain model. Both are $0$-form symmetries and act on local operators. The operators charged under the momentum (resp.~winding) symmetry are $e^{in\phi}$ (resp.~$e^{iw\tilde{\phi}'}$). Notice that the winding current is conserved because we eliminated the vortices by imposing $dn=0$. In the unmodified Villain action, this current is no longer conserved and we lose the winding symmetry. Finally notice that, consistently with its continuum limit, the term $\sum_l j^m_l (*\tilde{j}^w)_l $ generates the marginal operator which changes the radius of the compact boson.

Given the explicit expression for the two U(1) currents, we can show that they enjoy the same mixed 't Hooft anomaly of the continuous theory \cite{Gorantla:2021svj}. Indeed by coupling the theory to the two background $U(1)$ gauge fields $A^m$ and $A^w$, we get
\begin{equation}\label{actionbackgrounds}
    S = \frac{\beta}{2} \sum_l (d\phi -A^m- 2\pi n)^2_l +i\sum_p (\phi')_p( dn + M)_p + \frac{i}{2\pi} \sum_l A^{w}_l(d\phi -A^m - 2\pi n)_l + i \sum_v \phi_v M_v' \ ,
\end{equation}
with gauge transformations
\begin{equation}
    \begin{aligned}
    &\phi \rightarrow \phi + \lambda + 2\pi k_{\phi}\ ,  & \phi' \rightarrow \phi' + \lambda' + 2\pi k_{\phi'}\ , \\
    &n \rightarrow n + dk_{\phi} - q_m\ , &  A^{w} \rightarrow A^w + \delta \lambda' + 2\pi q_{w}\ ,\\
    & A^m \rightarrow A^m + d\lambda + 2\pi q_m\ , &M' \rightarrow M' + \delta q_w\ , \\
    & M \rightarrow M +dq_m\ .
    \end{aligned}
\end{equation}
The fields $M\in C^2(\Lambda,\bZ)$ and $M'\in C^0(\Lambda,\bZ)$ need to be introduced so that $A^m$ and $A^w$ can correctly be considered fields in $C^1(\Lambda,U(1))$. 
However it is easy to show that the action is not gauge invariant. In fact, under the above gauge transformations, we have 
\begin{equation}
    \delta S = -\frac{i}{2\pi} \sum_p \lambda_p' (dA^m -2\pi M)_p + i \sum_l q_w A^m + i\sum_v \lambda_v (M' + \delta q_w)_v\ .
\end{equation}
Note that we have also ignored all the terms that are in $2\pi i\bZ$, produced by sums over integer fields such as $n$, $M$, $M'$, $k_\phi$, $k_{\phi'}$, $q_m$, and $q_w$.
This variation cannot be canceled by adding local counterterms hence, it is a mixed anomaly between the winding and momentum symmetries. Indeed, if we set to zero either of the gauge fields and their corresponding gauge transformations, the anomaly disappears. Note that in order to see the anomaly, it is not necessary to implement the $U(1)$ periodicities of the gauge fields $A^m$ and $A^w$. Indeed, after setting to zero all $M$s and $q$s the anomaly persists as $-\frac{i}{2\pi} \sum_p \lambda_p' (dA^m)_p$.

The presence of the mixed anomaly can be traced to the fact that the gauge invariances mix degrees of freedom on the links and on the vertices. As a result, the winding current $j^w = - *n$ must be dressed with a $d\phi$ to make it gauge invariant under $k_{\phi}$. But this dressed current is not invariant under a background gauge transformation for the momentum symmetry.

Finally, the present Euclidean lattice model enjoys a T-dual formulation, that descends directly from the possibility to perform a Poisson resummation from the integer variables $n$ to some dual variables $k$ \cite{Sulejmanpasic:2019ytl, Gorantla:2021svj}. One first writes the action as
\be
S=\frac{\beta}{2}\sum_l(d\phi-2\pi n)_l^2+i\sum_p \phi'_p(dn)_p = \frac{\beta}{2}\sum_l(d\phi-2\pi n)_l^2+i\sum_l (\delta\phi')_ln_l\ .
\ee
The partition function is a sum over $n_l$ for every link $l$. Each of these sums can be reexpressed, using Poisson resummation, as a sum over a new integer variable $k_l$, yielding a new (but completely equivalent) action
\be
S=\frac{1}{8\pi^2\beta}\sum_l(\delta\phi'-2\pi k)_l^2-\frac{i}{2\pi}\sum_l (d\phi)_l (\delta\phi'-2\pi k)_l = \frac{1}{8\pi^2\beta}\sum_l(\delta\phi'-2\pi k)_l^2+i\sum_l (d\phi)_l k_l\ ,
\ee
where is the second equality we have eliminated an exact term. Finally, on the dual lattice the action reads
\begin{align}
    S= \frac{1}{8\pi^2\beta}\sum_{\tilde l}(d\tilde\phi'-2\pi \tilde k)_{\tilde l}^2+i\sum_{\tilde l} (\delta \tilde \phi)_{\tilde l} \tilde k_{\tilde l}\ ,
\end{align}
which has the same form as the original action, with a T-dual coupling $\tilde \beta=1/(4\pi^2\beta)$ (or equivalently $\tilde R=1/R)$.

\subsection{Modified Villain as a Topological Interface}

We want now to show that the modified Villain procedure which gives rise to a compact boson starting from a non-compact boson is indeed a \emph{topological manipulation} corresponding to the gauging of the $\bZ$ subgroup of the $\bR$ shift symmetry of the non-compact boson. To show this, we explicitly construct an interface separating this two theories, and show its topological nature.

Before delving into the argument of the topological nature of the procedure, let us show that we can undo the modified Villain procedure and get a non-compact boson starting from the compact one. Following the intuition coming from the continuum theory, we need to gauge the winding symmetry with flat connections.

To gauge the winding symmetry we couple a gauge field $a$ to the winding current in the modified Villain model:
\begin{equation} \label{XY:wind_gauge_lattice}
    S = \frac{\beta}{2}\sum_l(d\phi - 2\pi n)^2_l + i\sum_p(\phi')_p(dn)_p -i \sum_l a_ln_l \ ,
\end{equation}
with new gauge transformations acting on $a$ and on $\phi'$: 
\begin{equation} \label{XY:gauge_winding}
    \begin{aligned}
        & a \rightarrow a + 2\pi q_w + \delta \lambda' \\
        & \phi' \rightarrow \phi' + \lambda'\ ,
    \end{aligned}
\end{equation}
with ${q}_w \in C^1({\Lambda}_2,\bZ)$ and ${\lambda}' \in C^1({\Lambda}_2,U(1))$.\footnote{The coupling term in \eqref{XY:wind_gauge_lattice} can be seen as originating from a coupling on the dual lattice $i \sum_{\tilde{l}} \tilde{a}_{\tilde{l}} (*n)_{\tilde{l}} = -i \sum_l a_ln_l$,
where we have used that $l = *\tilde{l}$ in two dimensions. We also label $\tilde a =*a$, and note that closedness (i.e.~flatness) of $\tilde a$, $d\tilde a=0$, translates to co-closedness of $a$, $\delta a=0$.} Under the $\lambda'$ gauge transformation, this coupling term transforms as
\begin{equation}
    -i\sum_l (\delta\lambda')_l n_l = -i\sum_p \lambda'_p (dn)_p
\end{equation}
and indeed cancels the gauge transformation of $\phi'$. The ${q}_w$ gauge transformation ensures ${a}$ is valued in $U(1)$. As pointed out earlier, the current $j^w=-*n$ is not gauge invariant (under $n \rightarrow n + dk_\phi$) and only the combination $\frac{1}{2\pi}*(d\phi - 2\pi n)$ is. But the above coupling with $a$ is gauge invariant when the latter is a flat gauge field, as can be seen by integrating by parts (given that we sum over all lattice links in the coupling term). We thus need to impose $\delta a = 0$. Notice in passing that because of the flatness of $a$, it would also be redundant to add a field like $M'$ in \eqref{actionbackgrounds}, since we must already have $\delta q_w=0$.

We then need to impose $d\tilde{a} = 0$, or equivalently $\delta a=0$. Analogously to the continuum derivation, we enforce the flatness condition via a Lagrangian multiplier. But since $a$ is a $U(1)$ gauge field with gauge invariance \eqref{XY:gauge_winding}, the flatness condition $\delta a = 0$ must actually be interpreted as $\delta a = 2\pi \bZ$. Putting all together, we start with the following partition function
\begin{equation}
    Z = \frac{\left( \prod_v \int_{-\infty}^{\infty}d\phi_v\right)\left( \prod_p \int_{-\infty}^{\infty}\frac{d\phi_p'}{2\pi}\right)\left(\prod_l \sum_{n_l \in \mathbb{Z}}\right)\left( \prod_l \int_{-\infty}^{\infty} da_l \right) \left( \prod_v \sum_{m'_v \in \bZ} \delta((\delta a)_v -2\pi m'_v) \right)  }{\prod_v \text{Vol}(\bZ)_{(k_{\phi})_v} \prod_p \text{Vol}(\bZ)_{(k_{\phi'})_p} \prod_l \text{Vol}(\bZ)_{(q_w)_l} \prod_p \text{Vol}(U(1))_{\lambda'_p}} e^{-S}\,.
\end{equation}
The denominator corresponds to all the gauge redundancies (we denote $\text{Vol}(\bZ)_{k}=\sum_{k\in\bZ}$ and $\text{Vol}(U(1))_{\lambda}=\int_0^{2\pi}d\lambda$). It will turn out to be the most convenient normalization. We then use the Poisson resummation formula to convert the sum over Kronecker deltas into a sum of exponentials
\begin{equation} \label{XY:TOPO:constraint}
    \sum_{m'_v} \delta((\delta a)_v -2\pi m'_v) = \frac{1}{2\pi}\sum_{m_v}  e^{-i (\delta a)_v m_v}\,.
\end{equation}
Integrating by parts all these terms we obtain
\be \label{XY:TOPO:z}
     Z  = \frac{\left(\prod_v \int_{-\infty}^{\infty}d\phi_v\right)\left( \prod_p \int_{-\infty}^{\infty}\frac{d\phi_p'}{2\pi}\right)\left(\prod_l \sum_{n_l \in \mathbb{Z}}\right)\left( \prod_l \int_{-\infty}^{\infty} da_l \right) \left( \prod_v \frac{1}{2\pi}\sum_{m_v \in \bZ} \right)  }{\prod_v \text{Vol}(\bZ)_{(k_{\phi})_v} \prod_p \text{Vol}(\bZ)_{(k_{\phi'})_p}\prod_l \text{Vol}(\bZ)_{(q_w)_l} \prod_p \text{Vol}(U(1))_{\lambda'_p}}e^{-S'}\ , 
\ee
with
\be 
    S'=\frac{\beta}{2} \sum_l (d\phi - 2\pi n)^2_l + i \sum_p (\phi')_p(dn)_p - i \sum_l a_l n_l + i\sum_l a_l (dm)_l\ .
\ee
We next use the gauge invariance $\text{Vol}(\bZ)_{q_w}$ on $a$ to gauge fix it to a $2\pi$ interval. This allows us to integrate out each $a_l$ yielding Kronecker deltas
\begin{equation}
    \int_{0}^{2\pi}da_l e^{ia_l(n_l - (dm)_l)} = 2\pi \delta_{n_l, (dm)_l}\,.
\end{equation}
These constraints force $n$ to be exact. They thus cancel the non-trivial winding part of $n$ just as in the continuum where the non-trivial winding part of $\phi$ was removed by integrating out the gauge field. 

The sums over $n$ can then be performed straightforwardly, replacing $n=dm$ in the action. As a result $\phi'$ disappears from the action, and its integrals can be simplified against the volumes of its gauge transformations.
The partition function then becomes
\begin{equation}
    Z=\frac{\left(\prod_v \int_{-\infty}^{\infty}d\phi_v\right) \left( \prod_v \frac{1}{2\pi}\sum_{m_v \in \bZ} \right)(2\pi)^{L-P}}{\prod_v \text{Vol}(\bZ)_{(k_{\phi})_v} } e^{ -\frac{\beta}{2} \sum_l (d\phi - 2\pi dm)^2_l  }\ .
\end{equation}
We finally use the gauge invariance $k_{\phi}$
\begin{equation}
    \begin{aligned}
        &\phi \rightarrow \phi + 2\pi k_{\phi}\\
        &dm \rightarrow dm + dk_{\phi}
    \end{aligned}
\end{equation}
to set $dm = 0$. There is still a gauge transformation left parametrized by an exact $k_{\phi}$ i.e. $(k_{\phi})_v = k^0 $ constant over each connected component of the lattice. The same holds for $m$ (we only set $dm=0$ and so there is a sum over the $m$ constant over each connected component left). But these two cancel and we are left with
\begin{equation}
    Z =  \frac{1}{(2\pi)^{\chi}}\prod_v  \int_{-\infty}^{\infty} d\phi_v e^{-\frac{\beta}{2} \sum_l (d\phi )^2_l}\ ,
\end{equation}
with $\chi = V+P-L$, the Euler characteristic of the lattice. This is the partition function for the non-compact boson.

Until now we have shown that the non-compact and compact bosons are related by two topological manipulations: the modified Villain procedure, which corresponds to gauging a $\bZ$ subgroup of the shift symmetry, and the flat gauging of the winding symmetry. We now aim to show that one can indeed define a topological interface between the two theories.\footnote{The proof closely follows the analysis of \cite{Choi:2021kmx}, in which half-space gaugings of discrete symmetries were shown to produce topological interfaces in the modified Villain model.} 

To establish this, we consider a straight interface embedded within our square lattice, serving as a boundary between the compact and non-compact boson.
The two partition functions are :
\begin{itemize}
    \item[-] Non-compact boson: 
    \begin{equation}
        Z_{\text{nc}} = \prod_v \int_{-\infty}^{\infty} d\phi_ve^{-\frac{\beta}{2}\sum_l(d\phi)_l^2}      
    \end{equation}

\item[-] Compact boson (modified Villain formulation):
\begin{equation}
    Z_{\text{c}} = \frac{\prod_v \int_{-\infty}^{\infty} d\phi_v\prod_p \int_{-\pi}^{\pi}\frac{d{\phi'}_p}{2\pi}\prod_l \sum_{n_l \in \mathbb{Z}}}{\prod_v\text{Vol}(\mathbb{Z})_{k_v}}e^{-\frac{\beta}{2}\sum_l (d\phi - 2\pi n)^2_l + i\sum_p (\phi')_p(dn)_p},
\end{equation}
    with gauge invariance: 
\begin{equation}
    \begin{aligned}
        &\phi \rightarrow \phi + 2\pi k \\
        &n \rightarrow n + dk \,.
    \end{aligned}
\end{equation}
\end{itemize}

\begin{figure}[t]
    \begin{center}
    \begin{tikzpicture}

\def\rows{5} 
\def\cols{5} 
\def\size{1} 
\def\markx{2} 
\def\marky{2} 
\def\shiftx{7} 

\foreach \i in {0,...,\rows} {
    \foreach \j in {0,...,\cols} {
        \fill (\i*\size, \j*\size) circle (2pt);

        \ifnum \i<\rows
            \draw[dotted] (\i*\size, \j*\size) -- ({(\i+1)*\size}, \j*\size);
        \fi

        \ifnum \j<\cols
            \ifnum \i=\markx
                \draw[red, thick] ({(\i)*\size}, \j*\size) -- ({(\i)*\size}, {(\j+1)*\size}); 
            \else
                \draw[dotted] (\i*\size, \j*\size) -- (\i*\size, {(\j+1)*\size}); 
            \fi
        \fi
    }
}

\node[] at (6,2.5) {$=$};

\foreach \i in {0,...,\rows} {
    \foreach \j in {0,...,\cols} {
        \fill (\i*\size + \shiftx, \j*\size) circle (2pt);

        \ifnum \i<\rows
            \draw[dotted] (\i*\size + \shiftx, \j*\size) -- ({(\i+1)*\size + \shiftx}, \j*\size);
        \fi

        \ifnum \j<\cols
            \ifnum \i=\markx
                \ifnum \j=\marky
                    \draw[black,dotted] ({(\i)*\size + \shiftx}, \j*\size) -- ({(\i)*\size + \shiftx}, {(\j+1)*\size}); 
                \else
                    \draw[blue, thick] ({(\i)*\size + \shiftx}, \j*\size) -- ({(\i)*\size + \shiftx}, {(\j+1)*\size}); 
                \fi
            \else
                \draw[dotted] (\i*\size + \shiftx, \j*\size) -- (\i*\size + \shiftx, {(\j+1)*\size}); 
            \fi
        \fi
    }
}

\draw[blue, thick] ({(\markx+1)*\size + \shiftx}, {(\marky)*\size}) -- ({(\markx+1)*\size + \shiftx}, {(\marky+1)*\size});
\draw[blue, thick] ({(\markx)*\size + \shiftx}, {(\marky)*\size}) -- ({(\markx+1)*\size + \shiftx}, {(\marky)*\size});
\draw[blue, thick] ({(\markx)*\size + \shiftx}, {(\marky+1)*\size}) -- ({(\markx+1)*\size + \shiftx}, {(\marky+1)*\size});

\node[] at (0.7,2.5) {\large{Non compact}};
\node[] at (5. + \shiftx,2.5) {\large{Compact}};

\node[] at (4.2,2.5) {\large{Compact}};
\node[] at (1. + \shiftx,2.5) {\large{Non compact}};

\node[] at (2.2,3.2) {1};
\node[] at (3.2,3.2) {2};
\node[] at (2.2,2.2) {4};
\node[] at (3.2,2.2) {3};

\node[] at (9.2,3.2) {1};
\node[] at (10.2,3.2) {2};
\node[] at (9.2,2.2) {4};
\node[] at (10.2,2.2) {3};

\end{tikzpicture}
    \end{center}
    \caption{Deforming the interface between non-compact and compact models. The red interface is $\Sigma$ and the blue one is $\Sigma'$. }
    \label{xy:topo:interface}
\end{figure}
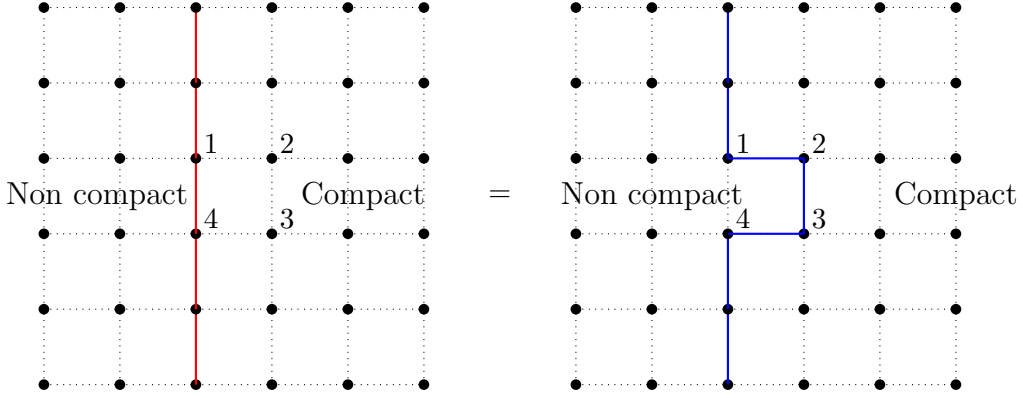

On the interface, we impose Dirichlet boundary conditions for the gauge field i.e. $ n_l  = 0 \quad \forall l\in \Sigma$, hence the $\phi_v$ degrees of freedom on the interface are non-compact. Deforming $\Sigma$ into $\Sigma'$ (see Figure \ref{xy:topo:interface}) amounts to:
\begin{itemize}
    \item[(1)] Integrate out $\phi_{1234}'$ which sets $(dn)_{1234} = 0 \Leftrightarrow n_{12} + n_{23} + n_{34} = 0$ 
    \item[(2)] Set $n_{12} = n_{23} = n_{34} = 0$ by gauge fixing 
    \item[(3)] Fix the gauge at vertices $2$ and $3$
\end{itemize}
 We now show that these three conditions allow a topological deformation of $\Sigma$ into $\Sigma'$ i.e. $Z(\Sigma) = Z(\Sigma')$. (1) sets $n_{23} = -n_{12} - n_{34}$ and (3) precisely allows (2). In fact, defining $Z$ as the partition function of all the lattice degrees of freedom except $\phi_3$, $\phi_2$, $n_{34}$, $n_{23}$, $n_{12}$ and $\phi_{1234}'$, we have:
 \begin{equation}
   Z(\Sigma) = Z \; \frac{\int_{-\infty}^{\infty}d\phi_2 \int_{-\infty}^{\infty}d\phi_3 \int_{0}^{2\pi}\frac{d\phi_{1234}'}{2\pi} \sum_{n_{34} \in \mathbb{Z}}\sum_{n_{23} \in \mathbb{Z}}\sum_{n_{12} \in \mathbb{Z}}}{\text{Vol}(\mathbb{Z})_{k_2} \text{Vol}(\mathbb{Z})_{k_3} } e^{-S}
 \end{equation}
Following (1), we integrate out $\phi_{1234}'$ which cancels the sum over $n_{23}$: 
\begin{equation}
     Z(\Sigma) = Z \; \frac{\int_{-\infty}^{\infty}d\phi_2 \int_{-\infty}^{\infty}d\phi_3  \sum_{n_{34} \in \mathbb{Z}}\sum_{n_{12} \in \mathbb{Z}}}{\text{Vol}(\mathbb{Z})_{k_2} \text{Vol}(\mathbb{Z})_{k_3} } e^{-S}
\end{equation}
Following (3) and (2), we set $n_{12} = n_{34} = 0$ using the two gauge invariances $k_2$ and $k_3$. We end up with:
\begin{equation}
     Z(\Sigma) = Z \; \int_{-\infty}^{\infty}d\phi_2 \int_{-\infty}^{\infty}d\phi_3 e^{-S} = Z(\Sigma')
\end{equation}
We see that the two gauge transformations precisely cancel the two sums over the integer gauge field $n$. The various normalization factors are needed for this strict equality, so this is an a posteriori justification. We conclude that deforming an interface between non-compact and compact bosons is a trivial (topological) operation because the two interfaces are gauge equivalent.

Additionally, this topological property of the interface is evident from the presentation of the model \eqref{eq: mvmodel_2} with $\R$ degrees of freedom at each vertex and residual integer degrees of freedom localized on the non-trivial $1$-cycles. These $1$-cycles can be deformed in a gauge invariant way, so that any deformation of an interface as described above can be made trivial after a convenient choice of positioning for the integer $1$-cycles.

\subsection{Non-Invertible T-duality Defect}
We would like to construct the non-invertible T-duality defect at a generic value $\beta$ in the modified Villain theory. To this aim, we split the lattice $\Lambda$ into a left part $\Lambda_L$, a right part $\Lambda_R$ and an interface $I$ (the latter is constituted only of vertices and links between them), and then we perform a concatenation of gaugings only on the right side of the interface, which send the coupling $\beta$ to itself in a non-trivial way. More specifically:
 \begin{enumerate}
     \item We gauge $U(1)^w$ with flat connections.
     \item We gauge $\Z \subset \R$ to obtain a compact theory with coupling $1/4\pi^2 \beta$.
     \item We perform T-duality to come back to the original presentation.
 \end{enumerate}
\paragraph{Gauging the winding symmetry.}Let us start with the first step. To gauge the winding symmetry with flat connections we consider the action
\begin{equation}
    S = \frac{\beta}{2} \sum_{l\in \Lambda_L \cup I \cup \Lambda_R}(d\phi - 2\pi n)_l^2 + i\sum_{p \in \Lambda_L \cup \Lambda_R} \phi_p'(dn)_p - i\sum_{l\in \Lambda_R}a_ln_l + i \sum_{v\in \Lambda_R} (\delta a)_v m_v\,.
\end{equation}
where we have introduced the minimal coupling to the current $-n_l$ and imposed flatness of $a_l$ via $m_v$.

Integrating over $a$ we trivialize all the $\bZ$ gauge fields $n_l$ on $\Lambda_R$. We end up with the action
\begin{equation}
    S = \frac{\beta}{2} \sum_{l\in \Lambda_L \cup I} (d\phi-2\pi n)^2_l + i\sum_{p \in \Lambda_L} \phi_p' (dn)_p + \frac{\beta}{2}\sum_{l \in \Lambda_R} (d\phi)_l^2- i\sum_{l\in I}\phi'_{p_l} n_l\,,
\end{equation}
where $p_l$ stands for the plaquette on $\Lambda_R$ next to the link $l \in I$. We notice that the variable $\phi'_{p_l}$ appears linearly and can be easily integrated out. Therefore we arrive to the action
\be
 S = \frac{\beta}{2} \sum_{l\in \Lambda_L } (d\phi-2\pi n)^2_l + i\sum_{p \in \Lambda_L} \phi_p' (dn)_p + \frac{\beta}{2}\sum_{l \in I\cup\Lambda_R} (d\phi)_l^2\ ,
\ee
where now $\phi_{v\in I}$ are decompactified, i.e. they have trivial gauge transformations. Importantly, while the field $\phi_v$ is defined on the entire lattice and it is smooth through the interface, the gauge field $n$ now lives only on $\Lambda_L$ and the integration over $\phi'$ imposed Dirichlet boundary conditions for it on $I$, i.e. we have 
\be
n_{l\in I}=0\,.
\ee
\paragraph{Gauging the $\bZ$ symmetry.} We can now compactify again the real boson on $\Lambda_R$. To choose the radius of the correct compactification it is convenient to redefine $\phi_{v \in \Lambda_R}$ as
\be
\phi_v \rightarrow \psi_v = 2\pi \beta\, \phi_v \,,
\ee
to get the new action
\begin{equation}
    S = \frac{\beta}{2} \sum_{l \in \Lambda_L}(d\phi - 2\pi n)_l^2 + i\sum_{p \in \Lambda_L} \phi_p' (dn)_p + \frac{\beta}{2}\sum_{l \in I} (d\phi)_l^2+  \frac{1}{8\pi^2 \beta}\sum_{l \in\Lambda_R} (d\psi)_l^2 \,.
\end{equation}
We now gauge the $\bZ\subset \bR$ shift symmetry by introducing the gauge field $m' \in C^1(\Lambda_R,\bZ)$ on the links and write the action
\begin{equation}
\begin{aligned}
     S =& \frac{\beta}{2} \sum_{l \in \Lambda_L }(d\phi - 2\pi n)_l^2 + i\sum_{p \in \Lambda_L} \phi_p' (dn)_p + \frac{\beta}{2}\sum_{l \in I} (d\phi)_l^2   \\
     &+  \frac{1}{8\pi^2 \beta}\sum_{l \in \Lambda_R} (d\psi - 2 \pi m')_l^2 + i \sum_{p\in \Lambda_R} \psi_p' (dm')_p 
\end{aligned}
\end{equation}
Again, since $m'$ is defined on $\Lambda_R$ we have to impose boundary conditions on $I$ and we choose Dirichlet b.c. $m'_{l\in I}=0$ to ensure that the interface remains topological.
\paragraph{T-duality and final result.} We can now perform T-duality on $\Lambda_R$ to come back to the original theory. This is done via the Poisson resummation on every link $l\in \Lambda_R$. We get
\begin{equation}
\begin{aligned}
     S =& \frac{\beta}{2} \sum_{l \in \Lambda_L }(d\phi - 2\pi n)_l^2 + i\sum_{p \in \Lambda_L} \phi_p' (dn)_p + \frac{\beta}{2}\sum_{l \in I} (d\phi)_l^2   \\
& +\frac{\beta}{2} \sum_{l \in \Lambda_R }(\delta \psi' - 2\pi k)_l^2  - \frac{i}{2\pi} \sum_{l\in \Lambda_R} (d\psi)_l(\delta \psi' - 2\pi k)_l  \,.
\end{aligned}
\end{equation}
Notice that since $m'_{l\in I}=0$ we have also $k_{l\in I}=0$.

To get the correct action on the right-hand-side of $I$ we need to map $\Lambda_R$ to its dual lattice. To do this we first rewrite
\begin{equation}
    \begin{aligned}
        -\frac{i}{2\pi} \sum_{l\in \Lambda_R} (d\psi)_l(\delta \psi' - 2\pi k)_l &= -\frac{i}{2\pi} \sum_{l\in \Lambda_R} (d\psi)_l (\delta \psi')_l + i\sum_{l}(d\psi)_l k_l  \\
        &=  \frac{i}{2\pi} \sum_{v \in I} \psi_v (\delta \psi')_{l_v} -i \sum_{v \in I} \psi_v k_{l_v}- i\sum_{v \in \Lambda_R}  \psi_v (\delta k)_v \\
        &= i\beta \sum_{v \in I} \phi_v (\delta \psi')_{l_v} - 2\pi i \beta \sum_{v \in I} \phi_v k_{l_v} - i\sum_{v \in \Lambda_R}  \psi_v (\delta k)_v
    \end{aligned}
\end{equation}
with $l_v$ for $v\in I$, the link next to $v$ in $\Lambda_R$ and where we used $\psi_v = 2\pi \beta \phi_v$ for $v \in I$. Evaluating on the dual lattice yields the result
\begin{equation}
\begin{aligned}
     S =& \frac{\beta}{2} \sum_{l \in \Lambda_L }(d\phi - 2\pi n)_l^2 + i\sum_{p \in \Lambda_L} \phi_p' (dn)_p+ \frac{\beta}{2}\sum_{l \in I} (d\phi)_l^2 \\
     & + i\beta \sum_{v \in I} \phi_v (\delta \psi'-2\pi k)_{l_v}  \\
     &+ \frac{\beta}{2}\sum_{\tilde{l} \in \tilde{\Lambda}_R} (d \tilde{\psi}' - 2 \pi \tilde{k})_{\tilde{l}}^2 + 
     i\sum_{\tilde{p} \in \tilde{\Lambda}_R}  \tilde{\psi}_{\tilde{p}} (d \tilde{k})_{\tilde{p}}\,,
\end{aligned}
\end{equation}
where we have defined $\tilde \psi'_{\tilde v}=* \psi'_p$, $\tilde k_{\tilde l}=*k_l$ and $\tilde \psi_{\tilde p}=*\psi_v$. The dual lattice $\tilde \Lambda_R$ on the right has a boundary $\tilde I$, that we want eventually to identify with $I$. On its vertices, we therefore have two degrees of freedom, $\phi$ and $\tilde\psi'\equiv \phi_R$, while on its links we have only $\tilde k \equiv n_R$, namely the gauge fields that make $\phi_R$ compact. Indeed, due to the boundary condition $n_{l \in I}=0$, $\phi$ is non-compact. On the left part of the lattice $\Lambda_L$ we also relabel all the fields $\phi_L$, $n_L$ and $\phi'_L$. Using these identifications we can rewrite the action as
\be
S=S_L+S_R+S_I\ ,
\ee
with
\bea
S_L&= \frac{\beta}{2} \sum_{l \in \Lambda_L\cup I}(d\phi_L - 2\pi n_L)_l^2 + i\sum_{p \in \Lambda_L} (\phi'_L)_p (dn_L)_p \ , \\
S_R&= \frac{\beta}{2} \sum_{l \in \tilde\Lambda_R}(d\phi_R - 2\pi n_R)_l^2 + i\sum_{p \in\tilde \Lambda_R} (\phi'_R)_p (dn_R)_p \ , \\
S_I &=i\beta \sum_{v \in I} \phi_v (d\phi_R-2\pi n_R)_{l_v} \,,
\eea
where in the interface action, $l_v$ is the link below the vertex $v$ (by convention), and the matching conditions from the left side are
\be
(\phi_L)_{v\in I} = \phi_v\ , \qquad (n_L)_{l\in I}=0\ .
\ee
Notice that in particular, the scalar field $\phi_{v \in I}$ does not have gauge transformations and the interface action is hence gauge invariant for any choice of $\beta$.\footnote{The infinite quantum dimension of this topological defect is a consequence of the fact that the $\mathbb{Z}$ gauge symmetry, and the corresponding gauge field $n_L$, is turned off along the interface. Since these gauge transformations are absent on $I$, one no longer divides by the corresponding gauge volume in the path integral and the interface supports a non-compact zero mode $\phi_v$. Thus the partition function diverges, similarly to what happens in the continuum description of the defect.} 

\begin{figure}[t]
    \begin{center}
    \begin{tikzpicture}[scale=1.5]

\def\rows{5}      
\def\cols{5}      
\def\size{1}      
\def\markx{2}     
\def\dualgap{0.5} 

\foreach \i in {0,...,\markx} {
    \foreach \j in {0,...,\cols} {
        \ifnum\i=\markx
            \fill[red] (\i*\size,\j*\size) circle (2pt);
        \else
            \fill (\i*\size,\j*\size) circle (2pt);
        \fi

        \ifnum\i<\markx
            \draw[dotted] (\i*\size,\j*\size) -- ({(\i+1)*\size},\j*\size);
        \fi

        \ifnum\j<\cols
            \ifnum\i=\markx
                \draw[red,dotted] (\i*\size,\j*\size) -- (\i*\size,{(\j+1)*\size});
            \else
                \draw[dotted] (\i*\size,\j*\size) -- (\i*\size,{(\j+1)*\size});
            \fi
        \fi
    }
}

\foreach \i in {0,...,\numexpr\cols-\markx-1\relax} {
    \foreach \j in {0,...,\numexpr\rows-1\relax} {
        \fill[green!60!black] ({(\markx+\dualgap+\i)*\size},{(\j+\dualgap)*\size}) circle (2pt);

        \ifnum\i<\numexpr\cols-\markx-1\relax
            \draw[green!60!black,dashed]
                ({(\markx+\dualgap+\i)*\size},{(\j+\dualgap)*\size})
                --
                ({(\markx+\dualgap+\i+1)*\size},{(\j+\dualgap)*\size});
        \fi

        \ifnum\j<\numexpr\rows-1\relax
            \draw[green!60!black,dashed]
                ({(\markx+\dualgap+\i)*\size},{(\j+\dualgap)*\size})
                --
                ({(\markx+\dualgap+\i)*\size},{(\j+1+\dualgap)*\size});
        \fi
    }
}

\foreach \i in {0,...,\numexpr\cols-\markx-1\relax} {
    \draw[green!60!black,dashed]
        ({(\markx+\dualgap+\i)*\size},{-0.5*\size})
        --
        ({(\markx+\dualgap+\i)*\size},{0.5*\size});

    \draw[green!60!black,dashed]
        ({(\markx+\dualgap+\i)*\size},{(\cols-0.5)*\size})
        --
        ({(\markx+\dualgap+\i)*\size},{(\cols+0.5)*\size});
}

\foreach \j in {0,...,\numexpr\rows-1\relax} {
    \draw[green!60!black,dashed]
        ({(\cols-0.5)*\size},{(\j+\dualgap)*\size})
        --
        ({(\cols+0.5)*\size},{(\j+\dualgap)*\size});
}

\foreach \i in {\numexpr\markx+1\relax,...,\cols} {
    \foreach \j in {0,...,\cols} {
        \fill (\i*\size,\j*\size) circle (2pt);

        \ifnum\i<\cols
            \draw[dotted] (\i*\size,\j*\size) -- ({(\i+1)*\size},\j*\size);
        \fi

        \ifnum\j<\cols
            \draw[dotted] (\i*\size,\j*\size) -- (\i*\size,{(\j+1)*\size});
        \fi
    }
}

\foreach \j in {0,...,\cols} {
    \draw[dotted] (\markx*\size,\j*\size) -- ({(\markx+1)*\size},\j*\size);
}

\foreach \j in {0,...,\numexpr\rows-1\relax} {
    \draw[-stealth, green!50!black, line width=0.6pt, shorten >=3pt, shorten <=3pt]
        ({(\markx+\dualgap)*\size},{(\j+\dualgap)*\size})
        to[out=150,in=-35] (\markx*\size,{(\j+1)*\size});
}

\node[red] at (\markx*\size,{(\cols+0.6)*\size}) {\LARGE$I$};

\foreach \j in {0,...,\cols} {
    \node[red, left] at ({\markx*\size-0.0005},{\j*\size+0.25}) {\large$\phi_v$};
}

\foreach \j in {0,...,\numexpr\rows-1\relax} {
    \node[green!60!black, right] at ({(\markx+\dualgap)*\size+0.0},{(\j+\dualgap)*\size+0.25}) {\large$\tilde{\psi}'_{\tilde v}$};
}

\end{tikzpicture}
    \end{center}
    \caption{The gluing of the left lattice $\Lambda_L$ to the dual right lattice $\tilde \Lambda_R$, along the interface I.}
\end{figure}
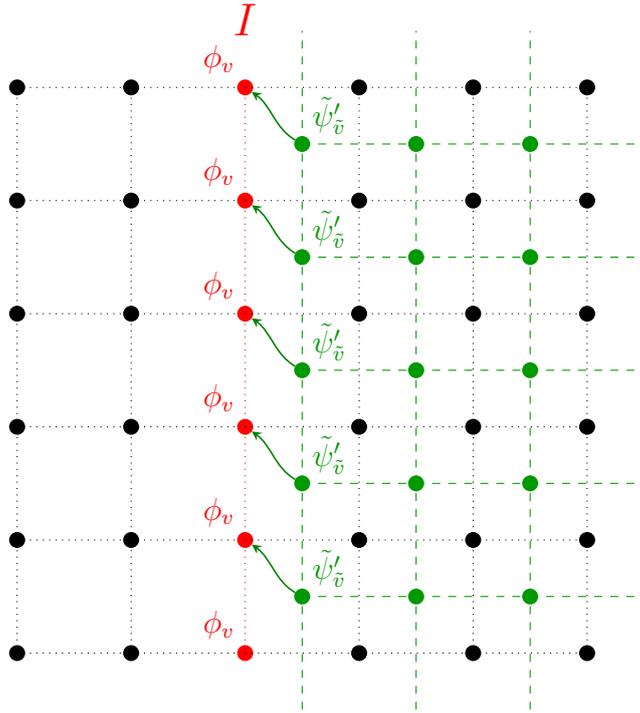

The interface action that we have obtained on the lattice is best compared to the one obtained in the continuum in the expression \eqref{tdualitydefectshort}, where the gauge invariance at any radius is also achieved by coupling the (compact) degrees of freedom on the right to some non-compact degrees of freedom defined on the defect, under the condition that the latter are to be matched with the compact degrees of freedom on the left. This is exactly as above.

Exactly as in the continuum, also in this case one can construct a minimal defect when $\beta = \frac{p}{2\pi q}$. Indeed we see that in these case we can relax the condition $(n_L)_{l\in I} = 0$ to be $(n_L)_{l\in I} \in q\bZ$, and reintroducing the gauge transformations
\be\label{eq: reduced gauge tranf}
\phi_{v\in I} \rightarrow \phi_{v\in I}  + 2\pi q k_{v \in I} \quad,\quad (n_L)_{l\in I} \rightarrow (n_L)_{l\in I} + q (dk)_{l\in I}
\ee
which make $\phi_{v \in I}$ compact. In order to make the interface action gauge invariant we now need to add an extra term (which is allowed as long as $n_L \not= 0$), getting
\be
S_{I,\text{min.}}^{\beta=\frac{p}{2\pi q}}= i \beta \sum_{v \in I}[ \phi_v(d\phi_R-2\pi n_R)_{l_v} +2\pi (n_L)_{l'_v} (\phi_R)_{v}]\,,
\ee
with $l'_v$ the link in $I$ above $v$.  
We notice that in the special case $q=1$ this action reproduces the result of \cite{Choi:2021kmx}.

\section{Hamiltonian Lattice Regularization}
\label{sec:hamiltonianlattice}
We now discuss the compact scalar model on a spatial lattice, namely a circular chain of vertices and links, with continuous time evolution governed by a simple Hamiltonian. As for the Euclidean lattice, we will see that compactness of the scalar must be obtained by a modified Villain prescription if one is to retain a winding symmetry. Once this is achieved, the model enjoys a T-dual formulation, and moreover T-duality can be promoted to a symmetry for any radius.

Let us start with a model for non-compact scalars (in this context, they could be called phonons) on a circular lattice with $L$ sites:\footnote{We take the lattice to be circular for definiteness.}
\be \label{VILL:PHONON}
    H_{\R} = \frac{\pi}{R^2} \sum_j p_j^2 + \frac{R^2}{4\pi} \sum_j (\phi_j - \phi_{j+1})^2\ ,
\ee
where we have chosen the couplings in view of the compactification that we are going to perform below.
Each $\phi_j$ is a real boson with momentum $p_j$ located on the site $j$. The commutation relations are thus $[\phi_i,p_j] = i\delta_{ij}$. Periodicity of the lattice is given by $\phi_{i+L} = \phi_i$. 

The above Hamiltonian has a $\R$ symmetry shifting all the bosons by some $\lambda \in \R$
\be
    \phi_i \rightarrow \phi_i + \lambda\ .
\ee
This symmetry is generated by the momentum charge
\be \label{HAM:Q^M:PHONON}
    Q^m = \sum_{j=1}^L p_j\ ,
\ee
which indeed commutes with the Hamiltonian and is hence conserved.

By gauging a $\Z$ subgroup of the $\R$ symmetry of the above model, we obtain the Villain model \cite{Fazza:2022fss,Cheng:2022sgb}. The $\Z$ gauging is perfomed as follows. 
We first define integer gauge fields by adding on each link a conjugate pair of degrees of freedom such that $[p^n_{i+1/2},n_{j+1/2}]=i\delta_{ij}$, and imposing that $G_{i+1/2}\equiv e^{2\pi in_{i+1/2}}$ acts trivially. Secondly, we enforce the following gauge invariance
\be \label{VILL:gauge}
\begin{aligned}
    &\phi_i \rightarrow \phi_i + 2 \pi k_i \\
    & n_{i+1/2} \rightarrow n_{i+1/2} +k_i-k_{i+1}\ ,
\end{aligned}
\ee
with $k_i$ valued in $\Z$. The gauge invariance described above can be implemented by requiring that the operators generating the corresponding gauge transformations,
\begin{equation}
    G_i \equiv e^{\,i\left(p^n_{i-1/2} - p^n_{i+1/2} + 2\pi p_i\right)},
\end{equation}
act trivially on the physical Hilbert space. Thirdly, we naturally consider the gauge invariant Hamiltonian
\be \label{HAM:VILLAIN}
    H_V = \frac{\pi}{R^2} \sum_j p_j^2 + \frac{R^2}{4\pi} \sum_j (\phi_j - \phi_{j+1} - 2\pi n_{j+1/2})^2 + \ldots,
\ee
where the $\ldots$ refer to other terms possibly involving $p^n_{i+1/2}$'s.

Lastly, in order to properly implement the modified Villain prescription, we need to work with a flat gauge field as in the Euclidean modified Villain model. In the present context, the equivalent of the flatness condition in two dimensions becomes a condition of non-dynamicity of the gauge field in the Hamiltonian Villain model.
The Hamiltonian in \eqref{HAM:VILLAIN} already fulfills that condition if we remove the possible terms involving the $p^n_{i+1/2}$'s. 

The modified Villain Hamiltonian we will consider is therefore given by
\be \label{HAM:MOD_VILLAIN}
    H_{mV} = \frac{\pi}{R^2} \sum_j p_j^2 + \frac{R^2}{4\pi} \sum_j (\phi_j - \phi_{j+1} - 2\pi n_{j+1/2})^2\ .
\ee

\subsection{Symmetries and Anomalies}\label{sec: ham sym}
We now discuss the symmetries of the modified Villain model. We first still have a momentum symmetry from the phonon model \eqref{HAM:Q^M:PHONON} but which is reduced to $\R/\Z \cong U(1)^m$ after the gauging procedure. 

In addition to the momentum symmetry, we have a winding symmetry generated by
\be
    Q^w=\sum_j \frac{1}{2\pi}(\phi_j-\phi_{j+1}-2\pi n_{j+1/2})\ ,
\ee
which comes from the fact that each $n_{i+1/2}$ commutes with the Hamiltonian. However, the $n_{i+1/2}$'s are not gauge invariant, while the above combination is conserved and gauge invariant (i.e. commutes with $H_{mV}$ and all the $G_i$s). Indeed, we wrote $Q^w$ with the $\phi_i$s even though they will cancel after the sum so as to highlight that $Q^w$ is made of local currents
\be
    j^w_{i+1/2} = \frac{1}{2\pi}(\phi_j-\phi_{j+1}-2\pi n_{j+1/2})\ .
\ee
The internal symmetry group is hence $G=U(1)^m \times U(1)^w$, the same as the one of the continuous compact boson.\footnote{More precisely, in all versions of the model the full symmetry is $G=(U(1)^m \times U(1)^w)\rtimes Z_2$, with $\Z_2$ coming from reflection symmetry.} A quick way \cite{Cheng:2022sgb} to see that there is a mixed anomaly between $U(1)^m$ and $U(1)^w$ is to evaluate the commutator of the two currents
\be
[j^m_i, j^w_j]=[p_i,\frac{1}{2\pi}(\phi_j-\phi_{j+1}-2\pi n_{j+1/2})]= \frac{1}{2\pi}(-i\delta_{ij}+i\delta_{i,j+1})\neq 0\ .
\ee
We can trace the anomaly to the terms that one has to add to the winding current to make it gauge invariant. Note that in the continuum and on the Euclidean lattice, it is usually simpler to find the anomaly by turning on background gauge fields, but it also gives rise to a non-trivial mixed two-point function of the currents.

Given the momentum and winding currents, we can construct the corresponding symmetry defects which extend along time \cite{Cheng:2022sgb}. The defect of the winding symmetry placed at site $I$ can be obtained by modifying the Gauss law at site $I$ to be $G_I = e^{i \lambda'}\in U(1)^w$.  Due to the definition of $G_I$ we can redefine the momentum variable $p_I \rightarrow p_I +\lambda'/2\pi$ to re-obtain the standard Gauss law $G_I =1$, at the price of obtaining a locally-modified Hamiltonian
\be\label{eq:winding defect}
H^{(I)}_w = \frac{\pi}{R^2} \sum_{j\not=I} p_j^2 + \frac{\pi}{R^2}(p_I+\frac{\lambda'}{2\pi})^2+\frac{R^2}{4\pi} \sum_j (\phi_j - \phi_{j+1} - 2\pi n_{j+1/2})^2\ .
\ee
Similarly, the momentum defect can be constructed by shifting $n_{I+1/2} \rightarrow n_{I+1/2}+\lambda/2\pi$, obtaining a locally-modified Hamiltonian
\be\label{eq:mom defect}
H^{(I)}_m = \frac{\pi}{R^2} \sum_{j} p_j^2 +\frac{R^2}{4\pi} \sum_{j\not=I} (\phi_j - \phi_{j+1} - 2\pi n_{j+1/2})^2 + \frac{R^2}{4\pi} (\phi_I - \phi_{I+1} - 2\pi n_{I+1/2} -\lambda)^2\ .
\ee
The topological nature of these interfaces can be seen from the fact that one can identify $H^{(I)}_{m,w}$ with $H^{(I+1)}_{m,w}$ with a redefinition of variables. For the momentum symmetry, this is just $\phi_{I+1}' = \phi_{I+1}+\lambda$. Since $\lambda \sim \lambda +2\pi$ this is an allowed change of variable that does not spoil any Gauss law. For the winding symmetry, we define $p_I = p'_I - \lambda'/2\pi$. However, in order to preserve the Gauss laws one needs to supplement this with the opposite transformations $p^n_{I+1/2}={p^n}'_{I+1/2}-\lambda'$ and $p_{I+1} = p'_{I+1}+\lambda'/2\pi$, thus moving the interface.\footnote{In \cite{Seifnashri:2026ema, Cheng:2022sgb} the topological nature of these interfaces is proven by showing that there exist unitary operators $U_{m,w}$ such that $U_{m,w}^{\dagger}H_{m,w}^{(I)}U_{m,w} = H_{m,w}^{(I+1)}$. Indeed this is equivalent to finding the correct change of variables connecting the two hamiltonians.}

\subsection{Compactification Interface}
Also in the present Hamiltonian framework we can consider an interface between the compact modified Villain model and the non-compact model that we begun with.

For the sake of clarity, let us start from the compact model \eqref{HAM:MOD_VILLAIN}. Decompactification is simply translated into removing the gauge fields $n_{j+1/2}$ from some link $j=I$ and then onwards.\footnote{Since we are only interested in the interface between the two models, we are going to work as if the chain was infinite and without any periodicity. Of course, in a closed chain it is straightforward to add a second (inverse) transition so that periodicity is maintained.} Note that this entails that the gauge transformations related to $n_{j+1/2}$ for $j\geq I$ must also be trivialized, namely we should set all $k_{j\geq I}=0$. This in turn implies that $\phi_I$ has no longer any gauge invariance, it is hence non-compact. The last compact degrees of freedom is $\phi_{I-1}$, with gauge invariance given by
\be
\begin{aligned}
    &\phi_{I-1} \rightarrow \phi_{I-1} + 2 \pi k_{I-1} \\
    & n_{I-1/2} \rightarrow n_{I-1/2} +k_{I-1}\ .
\end{aligned}
\ee
The Hamiltonian including this interface then reads
\be 
    H_{R|\infty}^{(I)} = \frac{\pi}{R^2} \sum_j p_j^2 + \frac{R^2}{4\pi} \sum_{j\leq I-1} (\phi_j - \phi_{j+1} - 2\pi n_{j+1/2})^2
    + \frac{R^2}{4\pi} \sum_{j\geq I} (\phi_j - \phi_{j+1})^2\ .
\ee
The gauge constraints are unchanged up to $G_{I-1/2}$, and they are all trivial from $G_I$ on.

We can actually show that such an interface is topological, i.e.~we can shift it by a lattice site just by a redefinition of variables. For instance, let us define the new variables
\be\label{redefshift}
\begin{aligned}
    &\phi_{I-1}'=\phi_{I-1}-2\pi n_{I-1/2}\ ,
    &n_{I-3/2}'=n_{I-3/2} + n_{I-1/2}\ .
\end{aligned}
\ee
It is immediate to realize that $\phi_{I-1}'$ has no gauge invariance, while $n_{I-3/2}'$ has the reduced one $n_{I-3/2}'\to n_{I-3/2}' + k_{I-2}$. The gauge invariance related to $k_{I-1}$ only affects $n_{I-1/2}$, and can be used to eliminate altogether this degree of freedom. The Hamiltonian finally reads
\be 
\begin{aligned}
    H_{R|\infty}^{(I)} = &\frac{\pi}{R^2} \sum_j p_j^2 + \frac{R^2}{4\pi} \sum_{j\leq I-3} (\phi_j - \phi_{j+1} - 2\pi n_{j+1/2})^2 \\ &
    + \frac{R^2}{4\pi}(\phi_{I-2}-\phi_{I-1}'-2\pi n_{I-3/2}')^2 + \frac{R^2}{4\pi}(\phi_{I-1}'-\phi_{I})^2
    + \frac{R^2}{4\pi} \sum_{j\geq I} (\phi_j - \phi_{j+1})^2 \equiv H_{R|\infty}^{(I-1)}\ ,
\end{aligned}
\ee
which is the same as before, but with the defect shifted from $I$ to $I-1$. One can similarly show that the constraints trivialize, namely $G_{I-1}$ becomes $e^{-ip^{n'}_{I-1/2}}$, with $p^{n'}_{I-1/2}$ the variable conjugate to $n_{I-1/2}$ after the redefinitions \eqref{redefshift}. This constraint together with $G_{I-1/2}$ reiterate that one can remove the canonical pair involving $n_{I-1/2}$. Shifting the defect in the other direction, from $I-1$ to $I$, can be performed introducing this trivial canonical pair, and performing the opposite redefinitions.

Since the coupling $R^2$ has no meaning for non-compact fields, we can rescale the latter as $R\phi_j=\bar \phi_j$ for $j\geq I$ (we go back at locating the defect at the site $I$), to get the Hamiltonian
\be 
\begin{aligned}
    H_{R|\infty}^{(I)} = &\frac{\pi}{R^2} \sum_{j\leq I-1} p_j^2 + \pi \sum_{j\geq I} \bar p_j^2 +\frac{R^2}{4\pi} \sum_{j\leq I-2} (\phi_j - \phi_{j+1} - 2\pi n_{j+1/2})^2 \\
    & +\frac{R^2}{4\pi}\Big(\phi_{I-1} - \frac{1}{R}\bar\phi_{I} - 2\pi n_{I-1/2}\Big)^2
    + \frac{1}{4\pi} \sum_{j\geq I} (\bar\phi_j - \bar\phi_{j+1})^2\ .    
\end{aligned}
\ee
We can further build an interface between the same model at two different radii $R$ and $R'$ as follows. We first write the Hamiltonian for an interface between the non-compact scalar on the left, and a compact one (at radius $R'$) on the right
\be 
\begin{aligned}
    H_{\infty|R'}^{(J)} = & \pi \sum_{j\leq J} \bar p_j^2 +\frac{\pi}{{R'}^2} \sum_{j\geq J+1} p_j^2+ \frac{1}{4\pi} \sum_{j\leq J-1} (\bar\phi_j - \bar\phi_{j+1})^2  \\
    & +\frac{{R'}^2}{4\pi}\Big(\frac{1}{R'}\bar\phi_{J} - \phi_{J+1} - 2\pi n_{J+1/2}\Big)^2+\frac{{R'}^2}{4\pi} \sum_{j\geq J+1} (\phi_j - \phi_{j+1} - 2\pi n_{j+1/2})^2
    \ .    
\end{aligned}
\ee
As long as $I\leq J$, we can put both interfaces on the same (infinite) chain. When $I=J$, we obtain an interface between two different radii
\be \label{HRRprime}
\begin{aligned}
    H_{R|R'}^{(I)} = &\frac{\pi}{R^2} \sum_{j\leq I-1} p_j^2 +\pi  \bar p_I^2 +\frac{\pi}{{R'}^2} \sum_{j\geq I+1} p_j^2\\
    &+\frac{R^2}{4\pi} \sum_{j\leq I-2} (\phi_j - \phi_{j+1} - 2\pi n_{j+1/2})^2  +\frac{R^2}{4\pi}\Big(\phi_{I-1} - \frac{1}{R}\bar\phi_{I} - 2\pi n_{I-1/2}\Big)^2 \\
    &  +\frac{{R'}^2}{4\pi}\Big(\frac{1}{R'}\bar\phi_{I} - \phi_{I+1} - 2\pi n_{I+1/2}\Big)^2 +\frac{{R'}^2}{4\pi} \sum_{j\geq I+1} (\phi_j - \phi_{j+1} - 2\pi n_{j+1/2})^2
    \ .     
\end{aligned}
\ee
We see that for an interface between two generic radii, we cannot avoid having a pair of non-compact degrees of freedom at the site $I$ of the interface. Accordingly, the gauge constraints are all the usual ones, except $G_I$ which is trivialized, and the gauge fields $n_{I\pm1/2}$ have the reduced gauge transformations $n_{I\pm1/2}\to n_{I\pm1/2}\mp k_{I\pm1}$. 

The non-invertible nature of the above interface can be seen as follows. Let us consider the momentum and winding defects discussed in Section \ref{sec: ham sym}. As explained before, such defects can be freely moved to the next site or link by a change of variables. In the presence of our interface, this means that the momentum or winding defect can be moved eventually to the site of the interface, or the link next to it. For the winding defect, it is easy to see that because of the non-compactness of $\bar\phi_I$, the shift $\lambda'$ can be removed by absorbing into a redefinition of $\bar p_I$. In other words, the interface can absorb any winding defect. For the momentum defect, it is slightly more subtle. When trying to pass it through the interface, one realizes that the periodicity of the shift $\lambda$ gets multiplied by $R/R'$. Hence, for generic irrational $R/R'$, there are no non-trivial momentum defects consistent with the interface.

Finally, as in the other presentations of the compact boson, when $R'/R = q/p$ we can construct a minimal interface with only compact edge modes. Indeed in this case we can redefine $\bar\phi_I = p R'\chi_I$ to get the action
\be 
\begin{aligned}
    H_{R|R'=\frac{q}{p}R}^{(I)} = &\frac{\pi}{R^2} \sum_{j\leq I-1} p_j^2 +\pi  \bar p_I^2 +\frac{\pi}{{R'}^2} \sum_{j\geq I+1} p_j^2\\
    &+\frac{R^2}{4\pi} \sum_{j\leq I-2} (\phi_j - \phi_{j+1} - 2\pi n_{j+1/2})^2  +\frac{R^2}{4\pi}\Big(\phi_{I-1} - q\chi_{I} - 2\pi n_{I-1/2}\Big)^2 \\
    &  +\frac{{R'}^2}{4\pi}\Big(p\chi_{I} - \phi_{I+1} - 2\pi n_{I+1/2}\Big)^2 +\frac{{R'}^2}{4\pi} \sum_{j\geq I+1} (\phi_j - \phi_{j+1} - 2\pi n_{j+1/2})^2
    \     ,
\end{aligned}
\ee
and subsequently reintroduce the gauge variable $k_{I}\in \bZ$, acting as
\be\label{eq: new gauge transf.}
 n_{I - 1/2}\rightarrow n_{I - 1/2} + k_{I - 1} -q k_{I}\quad,\quad n_{I + 1/2}\rightarrow n_{I + 1/2} + p k_{I } - k_{I+1}\quad,\quad \chi_I \rightarrow \chi_I + 2\pi k_I\, 
\ee
and its corresponding Gauss constraint
\be
G_I \equiv e^{i \left( q \,p^n_{I - 1/2}-p \,p^n_{I + 1/2}+2\pi \rho_I\right)} = 1\ ,
\ee
with $\rho_I=pR'\bar p_I$ the momentum conjugate to $\chi_I$. 
\subsection{T-duality Defect}
T-duality amounts to the fact that one can relabel all the degrees of freedom to rewrite the theory in a similar form, albeit with different couplings, and with an exchange of the roles of momentum and winding symmetries. In other words, we expect to exchange the roles of $p$ with $n$, and consequently of $\phi$ with $p^n$.

A change of variables that does this task, taking into account gauge invariance, is the following:
\be
  \begin{aligned}
       \tilde{p}_{i+1/2} &= -\frac{1}{2\pi}( \phi_{i} - \phi_{i+1}-2\pi n_{i+1/2}) \\
      \tilde{\phi}_{i+1/2} &=  p^n_{i+1/2}   \\
       \tilde{n}_{i} &= \frac{1}{2\pi}(p^n_{i-1/2} - p^n_{i+1/2} +2\pi p_i) \\
       p^{\tilde{n}}_i&=\phi_i \ .
  \end{aligned}
\ee
It is straightforward to check that the dual tilded variables have canonical commutation relations, exactly like the original untilded ones.

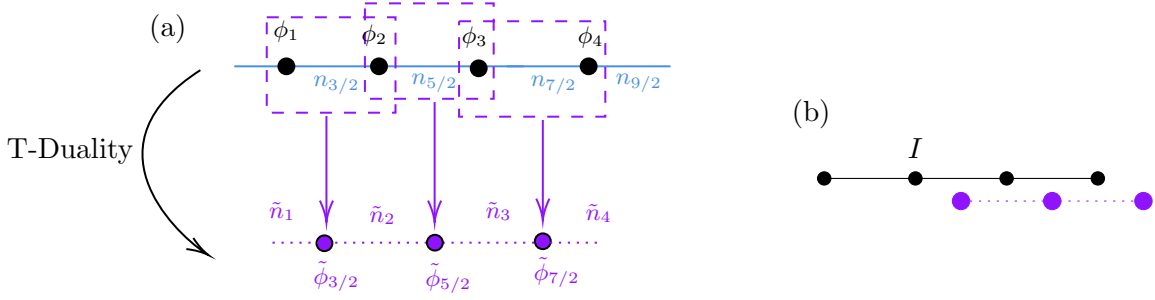
\begin{figure}[t]
\centering

\begin{minipage}[c]{0.48\textwidth}
\hspace*{5em}(a)\par
\vspace{-1.5em}
\centering

\tikzset{every picture/.style={line width=0.75pt}}        

\begin{tikzpicture}[x=0.75pt,y=0.75pt,yscale=-1,xscale=1]

\draw [color={rgb, 255:red, 74; green, 144; blue, 226 }  ,draw opacity=1 ]   (216,137) -- (360.47,137) ;
\draw  [fill={rgb, 255:red, 0; green, 0; blue, 0 }  ,fill opacity=1 ] (238.04,137) .. controls (238.04,134.81) and (239.81,133.04) .. (242,133.04) .. controls (244.19,133.04) and (245.96,134.81) .. (245.96,137) .. controls (245.96,139.19) and (244.19,140.96) .. (242,140.96) .. controls (239.81,140.96) and (238.04,139.19) .. (238.04,137) -- cycle ;
\draw  [fill={rgb, 255:red, 0; green, 0; blue, 0 }  ,fill opacity=1 ] (284.27,137) .. controls (284.27,134.81) and (286.05,133.04) .. (288.23,133.04) .. controls (290.42,133.04) and (292.19,134.81) .. (292.19,137) .. controls (292.19,139.19) and (290.42,140.96) .. (288.23,140.96) .. controls (286.05,140.96) and (284.27,139.19) .. (284.27,137) -- cycle ;
\draw [color={rgb, 255:red, 74; green, 144; blue, 226 }  ,draw opacity=1 ]   (352,137) -- (433.76,137) ;
\draw  [fill={rgb, 255:red, 0; green, 0; blue, 0 }  ,fill opacity=1 ] (334.04,138) .. controls (334.04,135.81) and (335.81,134.04) .. (338,134.04) .. controls (340.19,134.04) and (341.96,135.81) .. (341.96,138) .. controls (341.96,140.19) and (340.19,141.96) .. (338,141.96) .. controls (335.81,141.96) and (334.04,140.19) .. (334.04,138) -- cycle ;
\draw  [fill={rgb, 255:red, 0; green, 0; blue, 0 }  ,fill opacity=1 ] (388.92,137) .. controls (388.92,134.81) and (390.69,133.04) .. (392.88,133.04) .. controls (395.07,133.04) and (396.84,134.81) .. (396.84,137) .. controls (396.84,139.19) and (395.07,140.96) .. (392.88,140.96) .. controls (390.69,140.96) and (388.92,139.19) .. (388.92,137) -- cycle ;
\draw  [color={rgb, 255:red, 144; green, 19; blue, 254 }  ,draw opacity=1 ][dash pattern={on 4.5pt off 4.5pt}] (232.2,112.33) -- (296.47,112.33) -- (296.47,160.33) -- (232.2,160.33) -- cycle ;
\draw [color={rgb, 255:red, 144; green, 19; blue, 254 }  ,draw opacity=1 ]   (262,162) -- (262,213.49) ;
\draw [shift={(262,215.49)}, rotate = 270] [color={rgb, 255:red, 144; green, 19; blue, 254 }  ,draw opacity=1 ][line width=0.75]    (10.93,-3.29) .. controls (6.95,-1.4) and (3.31,-0.3) .. (0,0) .. controls (3.31,0.3) and (6.95,1.4) .. (10.93,3.29)   ;
\draw  [color={rgb, 255:red, 144; green, 19; blue, 254 }  ,draw opacity=1 ][dash pattern={on 4.5pt off 4.5pt}] (281.2,105.33) -- (345.47,105.33) -- (345.47,153.33) -- (281.2,153.33) -- cycle ;
\draw  [color={rgb, 255:red, 144; green, 19; blue, 254 }  ,draw opacity=1 ][dash pattern={on 4.5pt off 4.5pt}] (328.2,114.33) -- (401.2,114.33) -- (401.2,162.33) -- (328.2,162.33) -- cycle ;
\draw [color={rgb, 255:red, 144; green, 19; blue, 254 }  ,draw opacity=1 ]   (316,155) -- (316,213.49) ;
\draw [shift={(316,215.49)}, rotate = 270] [color={rgb, 255:red, 144; green, 19; blue, 254 }  ,draw opacity=1 ][line width=0.75]    (10.93,-3.29) .. controls (6.95,-1.4) and (3.31,-0.3) .. (0,0) .. controls (3.31,0.3) and (6.95,1.4) .. (10.93,3.29)   ;
\draw [color={rgb, 255:red, 144; green, 19; blue, 254 }  ,draw opacity=1 ]   (370,164) -- (370,212.49) ;
\draw [shift={(370,214.49)}, rotate = 270] [color={rgb, 255:red, 144; green, 19; blue, 254 }  ,draw opacity=1 ][line width=0.75]    (10.93,-3.29) .. controls (6.95,-1.4) and (3.31,-0.3) .. (0,0) .. controls (3.31,0.3) and (6.95,1.4) .. (10.93,3.29)   ;
\draw  [fill={rgb, 255:red, 144; green, 19; blue, 254 }  ,fill opacity=1 ] (366.04,225) .. controls (366.04,222.81) and (367.81,221.04) .. (370,221.04) .. controls (372.19,221.04) and (373.96,222.81) .. (373.96,225) .. controls (373.96,227.19) and (372.19,228.96) .. (370,228.96) .. controls (367.81,228.96) and (366.04,227.19) .. (366.04,225) -- cycle ;
\draw  [fill={rgb, 255:red, 144; green, 19; blue, 254 }  ,fill opacity=1 ] (257.04,226) .. controls (257.04,223.81) and (258.81,222.04) .. (261,222.04) .. controls (263.19,222.04) and (264.96,223.81) .. (264.96,226) .. controls (264.96,228.19) and (263.19,229.96) .. (261,229.96) .. controls (258.81,229.96) and (257.04,228.19) .. (257.04,226) -- cycle ;
\draw [color={rgb, 255:red, 144; green, 19; blue, 254 }  ,draw opacity=1 ] [dash pattern={on 0.84pt off 2.51pt}]  (235.09,225.71) -- (397.17,225.71) ;
\draw  [fill={rgb, 255:red, 144; green, 19; blue, 254 }  ,fill opacity=1 ] (312.17,225.71) .. controls (312.17,223.53) and (313.94,221.75) .. (316.13,221.75) .. controls (318.32,221.75) and (320.09,223.53) .. (320.09,225.71) .. controls (320.09,227.9) and (318.32,229.67) .. (316.13,229.67) .. controls (313.94,229.67) and (312.17,227.9) .. (312.17,225.71) -- cycle ;
\draw    (199.07,138.93) .. controls (141.84,175.79) and (184.34,219.24) .. (201.55,230.94) ;
\draw [shift={(203.07,231.93)}, rotate = 212.01] [color={rgb, 255:red, 0; green, 0; blue, 0 }  ][line width=0.75]    (10.93,-3.29) .. controls (6.95,-1.4) and (3.31,-0.3) .. (0,0) .. controls (3.31,0.3) and (6.95,1.4) .. (10.93,3.29)   ;

\draw (234,112.4) node [anchor=north west][inner sep=0.75pt]  [font=\footnotesize]  {$\phi _{1}$};
\draw (278,114.4) node [anchor=north west][inner sep=0.75pt]  [font=\footnotesize]  {$\phi _{2}$};
\draw (328,115.4) node [anchor=north west][inner sep=0.75pt]  [font=\footnotesize]  {$\phi _{3}$};
\draw (386,115.4) node [anchor=north west][inner sep=0.75pt]  [font=\footnotesize]  {$\phi _{4}$};
\draw (254,140.4) node [anchor=north west][inner sep=0.75pt]  [font=\footnotesize,color={rgb, 255:red, 74; green, 144; blue, 226 }  ,opacity=1 ]  {$n_{3/2}$};
\draw (302.96,139.4) node [anchor=north west][inner sep=0.75pt]  [font=\footnotesize,color={rgb, 255:red, 74; green, 144; blue, 226 }  ,opacity=1 ]  {$n_{5/2}$};
\draw (362.47,140.4) node [anchor=north west][inner sep=0.75pt]  [font=\footnotesize,color={rgb, 255:red, 74; green, 144; blue, 226 }  ,opacity=1 ]  {$n_{7/2}$};
\draw (405,139.4) node [anchor=north west][inner sep=0.75pt]  [font=\footnotesize,color={rgb, 255:red, 74; green, 144; blue, 226 }  ,opacity=1 ]  {$n_{9/2}$};
\draw (254,233.63) node [anchor=north west][inner sep=0.75pt]  [font=\footnotesize,color={rgb, 255:red, 144; green, 19; blue, 254 }  ,opacity=1 ]  {$\tilde{\phi }_{3/2}$};
\draw (310,234.63) node [anchor=north west][inner sep=0.75pt]  [font=\footnotesize,color={rgb, 255:red, 144; green, 19; blue, 254 }  ,opacity=1 ]  {$\tilde{\phi }_{5/2}$};
\draw (364,231.63) node [anchor=north west][inner sep=0.75pt]  [font=\footnotesize,color={rgb, 255:red, 144; green, 19; blue, 254 }  ,opacity=1 ]  {$\tilde{\phi }_{7/2}$};
\draw (232,203.63) node [anchor=north west][inner sep=0.75pt]  [font=\footnotesize,color={rgb, 255:red, 144; green, 19; blue, 254 }  ,opacity=1 ]  {$\tilde{n}_{1}$};
\draw (282,205.63) node [anchor=north west][inner sep=0.75pt]  [font=\footnotesize,color={rgb, 255:red, 144; green, 19; blue, 254 }  ,opacity=1 ]  {$\tilde{n}_{2}$};
\draw (340,203.63) node [anchor=north west][inner sep=0.75pt]  [font=\footnotesize,color={rgb, 255:red, 144; green, 19; blue, 254 }  ,opacity=1 ]  {$\tilde{n}_{3}$};
\draw (390,204.63) node [anchor=north west][inner sep=0.75pt]  [font=\footnotesize,color={rgb, 255:red, 144; green, 19; blue, 254 }  ,opacity=1 ]  {$\tilde{n}_{4}$};
\draw (244,158.93) node [anchor=north west][inner sep=0.75pt]    {$ \begin{array}{l}
\end{array}$};
\draw (101,171.53) node [anchor=north west][inner sep=0.75pt]   [align=left] {T-Duality};

\end{tikzpicture}
\end{minipage}
\hfill
\begin{minipage}[c]{0.48\textwidth}
\hspace*{4em}(b)\par
\vspace{-0.1em}
\centering

\begin{tikzpicture}[x=1cm,y=1cm]
\tikzset{
  site/.style={
    circle,
    draw=black,
    fill=black,
    minimum size=5pt,
    inner sep=0pt
  },
  dualsite/.style={
    circle,
    draw={rgb,255:red,144; green,19; blue,254},
    fill={rgb,255:red,144; green,19; blue,254},
    minimum size=6.5pt,
    inner sep=0pt
  },
  dualbond/.style={
    color={rgb,255:red,144; green,19; blue,254},
    dash pattern={on 0.84pt off 2.51pt}
  }
}

\coordinate (o1) at (0,0);
\coordinate (o2) at (1.2,0);
\coordinate (o3) at (2.4,0);
\coordinate (o4) at (3.6,0);

\draw (o1) -- (o4);
\foreach \p in {o1,o2,o3,o4}
  \node[site] at (\p) {};

\node[above=4pt] at (o2) {$I$};

\coordinate (d1) at (1.8,-0.3);
\coordinate (d2) at (3.0,-0.3);
\coordinate (d3) at (4.2,-0.3);

\draw[dualbond] (d1) -- (d3);
\foreach \p in {d1,d2,d3}
  \node[dualsite] at (\p) {};

\end{tikzpicture}
\end{minipage}

\caption{(a) T-duality map between the lattice and its dual. (b) How T-duality on half of the chain maps the original lattice into its dual.}
\label{fig:tduality}
\end{figure}

It is also interesting to observe how the gauge constraints $G_i$ and $G_{i+1/2}$ are expressed in the dual variables:
\be
\begin{aligned}
    &G_i =  e^{i(p^n_{i-1/2} - p^n_{i+1/2} + 2\pi p_i)} = e^{2\pi i \tilde n_i} \ , \\
    &G_{i+1/2}= e^{2\pi in_{i+1/2}} = e^{i( p^{\tilde{n}}_{i} - p^{\tilde{n}}_{i+1}+2\pi \tilde{p}_{i+1/2}) }\ .
\end{aligned}
\ee
They are also exactly exchanged. The mapping from one set of variables to the other can be roughly summarized as in Figure \ref{fig:tduality} (a).

Under this change of variables, the Hamiltonian of the modified Villain model becomes
\be
   H_{T} = \frac{1}{4\pi R^2} \sum_{j} (\tilde{\phi}_{j-1/2} - \tilde{\phi}_{j+1/2}-2\pi \tilde{n}_j)^2 + \pi R^2  \sum_j \tilde{p}_{j+1/2}^2 \ ,
  \ee 
which we recognize as the modified Villain model on the dual lattice with the coupling $\tilde R^2=1/R^2$. The self dual point is $R^2=1$.

We now discuss how to perform T-duality only on a part of the chain. Again, since we are only interested in the interface between the two T-dual descriptions, we are going to work as if the chain was infinite.

The way we operate the partial T-duality is as follows. We start with the original Hamiltonian, and then rewrite the terms after a given site $j=I$ in terms of the new, T-dual variables:
\be
\begin{aligned}
 H_T^{(I)} =& \frac{\pi}{R^2} \sum_{j\leq I} p_j^2 + \frac{1}{4\pi R^2} \sum_{j\geq I+1} (\tilde{\phi}_{j-1/2} - \tilde{\phi}_{j+1/2}-2\pi \tilde{n}_j)^2 \\ 
 &+ \frac{R^2}{4\pi} \sum_{j\leq I-1} (\phi_j - \phi_{j+1} - 2\pi n_{j+1/2})^2 + \pi R^2 \sum_{j\geq I} \tilde{p}_{j+1/2}^2\ .
\end{aligned}
\ee
More precisely, the original chain ends at the site $I$, which supports the degrees of freedom represented by $\phi_I$ and $p_I$. The link $I+1/2$ becomes a site of the new chain, with the degrees of freedom $p^n_{I+1/2}$ and $n_{I+1/2}$ traded for $\tilde\phi_{I+1/2}$ and $\tilde p_{I+1/2}$. Then the original site $I+1$ becomes a link in the new lattice, with degrees of freedom $p^{\tilde n}_{I+1}$ and $\tilde n_{I+1}$, and so on (see Figure \ref{fig:tduality} (b)). One can check that only these degrees of freedom appear in the Hamiltonian above. Also, since this operation is just a rewriting of the same Hamiltonian, it is clear that the site at which the rewriting starts can be moved right or left at no cost. The defect is hence topological.

Up to this point, it seems like we have a juxtaposition of two open chains, in original and dual variables respectively. What couples them is the fact that the gauge constraints around the site $I$ involve both original and dual variables. It is straightforward to see that the gauge constraints far enough from the site $I$, on both sides, will take the usual form in their respective variables. The gauge constraints that will couple original and dual variables are the ones at sites $I$ and $I+1/2$: 
    \be
    \begin{aligned}
        & G_{I} = e^{i(p^n_{I-1/2} - p^n_{I+1/2} +2\pi p_{I})} = e^{i(p^n_{I-1/2} -\tilde{\phi}_{I+1/2} +2\pi p_{I})} \ ,\\
        & {G}_{I+1/2} = e^{2\pi in_{I+1/2}} = e^{i(\phi_{I}-p^{\tilde{n}}_{I+1} +2\pi \tilde{p}_{I+1/2})}\ .
    \end{aligned}
    \ee
Another way to see this is that the definition of $\tilde{p}_{I+1/2}$ also involves $\phi_{I}$, so that we have a non-trivial commutation relation between the original and dual momenta
\be \label{VILL:T-DUAL:COMM}
     [\tilde{p}_{I+{1/2}},p_{I}]= -\frac{i}{2\pi} \ .
\ee
So the degrees of freedom at the (original) site $I$ and the ones at the (dual) site $I+1/2$ are not independent, though they still provide two canonical pairs. 

We can perform the fusion of two such defects. First of all the Hamiltonian with two defects at sites $I$ and $J$ is the following
\be
\begin{aligned}
 H_{T,T}^{(I,J)} =& \frac{\pi}{R^2} \sum_{j\leq I} p_j^2 + \pi R^2 \sum_{ I\leq j\leq J-1} \tilde{p}_{j+1/2}^2 +\frac{\pi}{R^2} \sum_{j\geq J} p_j^2\\ 
 &+ \frac{R^2}{4\pi} \sum_{j\leq I-1} (\phi_j - \phi_{j+1} - 2\pi n_{j+1/2})^2 + \frac{1}{4\pi R^2} \sum_{ I+1\leq j\leq J-1} (\tilde{\phi}_{j-1/2} - \tilde{\phi}_{j+1/2}-2\pi \tilde{n}_j)^2\\
 &+ \frac{R^2}{4\pi} \sum_{j\geq J} (\phi_j - \phi_{j+1} - 2\pi n_{j+1/2})^2\ .
\end{aligned}
\ee
We can then bring the two defects together by setting $J=I+1$. The Hamiltonian becomes
\be
\begin{aligned}
 H_{T^2}^{(I)} =& \frac{\pi}{R^2} \sum_{j} p_j^2 + \pi R^2 \tilde{p}_{I+1/2}^2 + \frac{R^2}{4\pi} \sum_{j\neq I} (\phi_j - \phi_{j+1} - 2\pi n_{j+1/2})^2 \ .
\end{aligned}
\ee
The gauge constraints are all as in the original model except the three following ones
\be
    \begin{aligned}
        & G_{I} =e^{i(p^n_{I-1/2} -\tilde{\phi}_{I+1/2} +2\pi p_{I})} \ ,\\
        & {G}_{I+1/2} =  e^{i(\phi_{I}-\phi_{I+1} +2\pi \tilde{p}_{I+1/2})}\ ,\\
        & G_{I+1} = e^{i(\tilde{\phi}_{I+1/2}-p^n_{I+3/2}+2\pi p_{I+1})}\ .
    \end{aligned}
\ee
The ${G}_{I+1/2}$ constraint together with the commutation relations $[\tilde{p}_{I+1/2},p_I]=-[\tilde{p}_{I+1/2},p_{I+1}]=-i/2\pi$ imply that we can define $n_{I+1/2}$ such that
\be
\tilde{p}_{I+1/2} = -\frac{1}{2\pi}( \phi_{I} - \phi_{I+1}-2\pi n_{I+1/2})\ .
\ee
The commutator $[\tilde{\phi}_{I+1/2},\tilde{p}_{I+1/2}]=i$ finally implies that $\tilde{\phi}_{I+1/2}=p^n_{I+1/2}$. The Hamiltonian  is thus perfectly equivalent to the original one, meaning that the fusion of two T-duality defects leaves behind the identity.

\subsection{The Non-Invertible Symmetry Defect}

The final step to write a T-duality symmetry defect is to insert at a site $I$ a defect implementing the transition from a chain at radius $R$ to a chain at radius $1/R$, then at some site $J>I$ (the inverse of) a T-duality defect as above, and then fuse them.

The Hamiltonian with the two defects at sites $I$ and $J$ respectively is given by
\be 
\begin{aligned}
    H_{R|1/R,T}^{(I,J)} = &\frac{\pi}{R^2} \sum_{j\leq I-1} p_j^2 +\pi  \bar p_I^2 +{\pi}{{R}^2} \sum_{ I+1\leq j\leq J} p_j^2+ \frac{\pi}{ R^2} \sum_{j\geq J} \tilde{p}_{j+1/2}^2\\
    &+\frac{R^2}{4\pi} \sum_{j\leq I-2} (\phi_j - \phi_{j+1} - 2\pi n_{j+1/2})^2  +\frac{R^2}{4\pi}\Big(\phi_{I-1} - \frac{1}{R}\bar\phi_{I} - 2\pi n_{I-1/2}\Big)^2 \\
    &  +\frac{1}{4\pi{R}^2}\Big(R\bar\phi_{I} - \phi_{I+1} - 2\pi n_{I+1/2}\Big)^2 +\frac{1}{4\pi{R}^2} \sum_{ I+1\leq j\leq J-1} (\phi_j - \phi_{j+1} - 2\pi n_{j+1/2})^2
    \\
    &  + \frac{R^2}{4\pi} \sum_{j\geq J+1} (\tilde{\phi}_{j-1/2} - \tilde{\phi}_{j+1/2}-2\pi \tilde{n}_j)^2  \ .
\end{aligned}
\ee
We can then set $J=I$, so that we finally have for the Hamiltonian with the symmetry defect
\be 
\begin{aligned}
    H_{T\mathrm{sym}}^{I} = &\frac{\pi}{R^2} \sum_{j\leq I-1} p_j^2 +\pi  \bar p_I^2 + \frac{\pi}{ R^2} \sum_{j\geq I} \tilde{p}_{j+1/2}^2\\
    &+\frac{R^2}{4\pi} \sum_{j\leq I-2} (\phi_j - \phi_{j+1} - 2\pi n_{j+1/2})^2  +\frac{R^2}{4\pi}\Big(\phi_{I-1} - \frac{1}{R}\bar\phi_{I} - 2\pi n_{I-1/2}\Big)^2 \\
    &  + \frac{R^2}{4\pi} \sum_{j\geq I+1} (\tilde{\phi}_{j-1/2} - \tilde{\phi}_{j+1/2}-2\pi \tilde{n}_j)^2  \ .
\end{aligned}
\ee
Note that as a part of the usual redefinition of variables that implements T-duality, we have to use the following one
\be
\tilde p_{I+1/2} = -\frac{1}{2\pi}(R\bar\phi_I-\phi_{I+1}-2\pi n_{I+1/2})\ .
\ee
In particular it implies
\be
[\tilde p_{I+1/2},\bar p_I]=-\frac{iR}{2\pi}\ ,
\ee
so that the degrees of freedom at sites $I$ and $I+1/2$ are still coupled. Finally, the gauge constraints around the interface are modified to be
\be
G_I=1\ , \qquad G_{I+1/2}=e^{i(R\bar \phi_I-p^{\tilde{n}}_{I+1} +2\pi \tilde{p}_{I+1/2})}\ .
\ee
Because of the permanence of the non-compact degree of freedom represented by $\bar\phi_I$, this defect implements a non-invertible symmetry. Indeed, in order to fuse two such defects, we just need to recall our previous results. It amounts to first fuse two T-duality defects, that yield the identity, and then two radius-changing defects. The result is the Hamiltonian \eqref{HRRprime}, with $R=R'$. Such a Hamiltonian still contains a (condensation) defect at the site $I$, which manifests itself by the presence of the non-compact degrees of freedom at that site.

Finally, one can also consider the special case $R^2 = p/q$ and construct a minimal T-duality defect with Hamiltonian
\be 
\begin{aligned}
    H_{T\mathrm{sym},R^2 = p/q}^{I} = &\frac{\pi}{R^2} \sum_{j\leq I-1} p_j^2 + \frac{\pi}{pq}\rho_I^2 + \frac{\pi}{ R^2} \sum_{j\geq I} \tilde{p}_{j+1/2}^2\\
    &+\frac{R^2}{4\pi} \sum_{j\leq I-2} (\phi_j - \phi_{j+1} - 2\pi n_{j+1/2})^2  +\frac{R^2}{4\pi}\Big(\phi_{I-1} - q\chi_{I} - 2\pi n_{I-1/2}\Big)^2 \\
    &  + \frac{R^2}{4\pi} \sum_{j\geq I+1} (\tilde{\phi}_{j-1/2} - \tilde{\phi}_{j+1/2}-2\pi \tilde{n}_j)^2  \ .
\end{aligned}
\ee
where $\chi_I$ is now compact with gauge transformations as \eqref{eq: new gauge transf.} (and corresponding Gauss laws) and the new commutation relation of its conjugate momentum is $[\rho_I,\tilde p_{I+1/2}]= \frac{ip}{2\pi}$.

Let us conclude by emphasizing the usefulness of the Hamiltonian perspective. The Hamiltonians introduced in this section allow us to determine the spectrum of the system explicitly. In the case of the modified Villain Hamiltonian \eqref{HAM:MOD_VILLAIN}, the spectrum computed in \cite{Cheng:2022sgb,Seifnashri:2026ema} can be used, as expected, to extract the scaling dimensions of the vertex operators of the continuum CFT. For the defect Hamiltonians analyzed in this work, instead, the corresponding spectrum determines the scaling dimensions of the states in the defect Hilbert space associated with the continuum interfaces and defects.

For irrational values of $R^2$, the presence of the non-compact edge mode $\bar{\phi}_I$ makes it manifest that the Hamiltonian admits eigenvalues labeled by a real parameter, and therefore has a continuous spectrum. This is the origin of the infinite quantum dimension expected for these interfaces. By contrast, the minimal interfaces and defects that arise at special values of the radii involve only compact edge modes, and their spectrum is therefore discrete.

\section*{Acknowledgments}

This research is funded through an ARC advanced project, and further supported by IISN-Belgium (convention 4.4503.15). R.A.~is a Research Director of the F.R.S.-FNRS (Belgium).  

\bibliographystyle{ytphys}
\baselineskip=0.85\baselineskip
\bibliography{references}

\end{document}